\documentclass[a4paper]{article}

\usepackage[utf8]{inputenc}
\usepackage[tt=false]{libertine}
\usepackage{microtype}
\usepackage{hyperref}
\usepackage{amsmath}
\usepackage{amssymb}
\usepackage{xspace}
\usepackage{xcolor}
\usepackage{graphicx}
\usepackage{booktabs}
\usepackage{setspace}
\usepackage[parfill]{parskip}
\usepackage[numbers,square,semicolon]{natbib}

\newcommand{\shortcite}[1]{\cite{#1}}
\renewcommand{\UrlFont}{\small\bf\ttfamily}

\newcommand{\plaintitle}{Computational Design with Crowds}
\newcommand{\plainauthors}{Yuki Koyama and Takeo Igarashi}
\newcommand{\plainkeywords}{computational design, crowdsourcing, user interface, Bayesian optimization, computer graphics}

\definecolor[named]{ACMPurple}{cmyk}{0.55,1,0,0.15}
\definecolor[named]{ACMDarkBlue}{cmyk}{1,0.58,0,0.21}

\hypersetup{colorlinks,
  pdftitle={\plaintitle},
  pdfauthor={\plainauthors},
  pdfkeywords={\plainkeywords},
  linkcolor=ACMPurple,
  citecolor=ACMPurple,
  urlcolor=ACMDarkBlue,
  filecolor=ACMDarkBlue}

\title{\plaintitle}
\author{Yuki Koyama\footnote{National Institute of Advanced Industrial Science and Technology (AIST)} \and Takeo Igarashi\footnote{The University of Tokyo}}
\date{}

\newcommand{\realnumber}{\mathbb{R}}

\newcommand{\bfx}{\mathbf{x}}
\newcommand{\bfy}{\mathbf{y}}

\newcommand{\calS}{\mathcal{S}}

\newcommand{\calX}{\mathcal{X}}
\newcommand{\argmax}{\mathop{\rm arg\,max}\limits}
\newcommand{\amax}[1]{\argmax_{#1}\,}

\newcommand{\eg}{\textit{e.g.,}\ }
\newcommand{\ie}{\textit{i.e.,}\ }
\newcommand{\cf}{\textit{c.f.,}\ }
\newcommand{\etal}{\textit{et al.}\@\xspace}

\newcommand{\figaesthetics}{
  \begin{figure*}[tb]
    \centering
    \includegraphics[width=\textwidth]{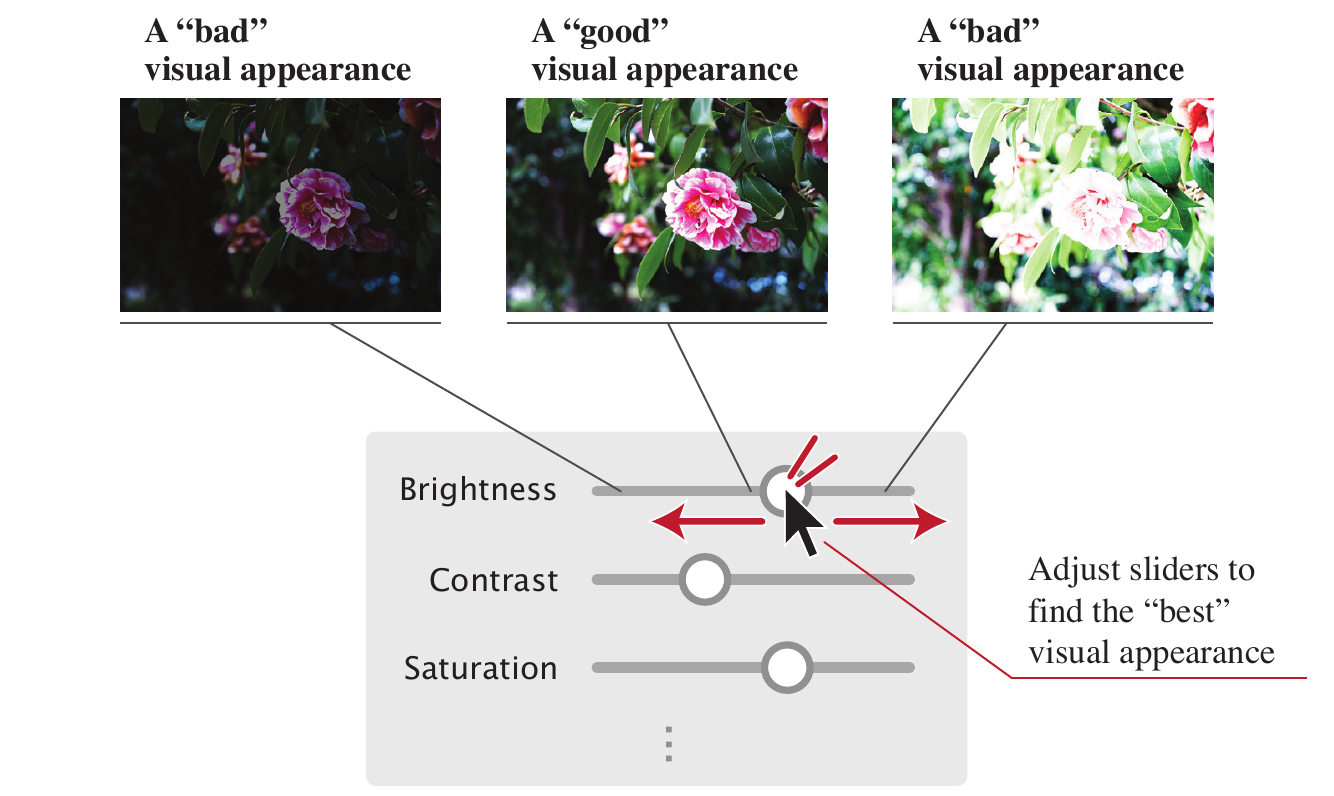}
    \caption[An example of design parameter tweaking, in which aesthetic preference is used as a criterion.]{
      \textbf{An example of design parameter tweaking, in which aesthetic preference is used as a criterion.}
      Photo color enhancement is one of such design scenarios, in which designers tweak sliders such as ``brightness'' so that they eventually find the parameter set that provides the best preferable photo enhancement.
    }
    \label{fig:aesthetics}
  \end{figure*}
}

\newcommand{\figsliders}{
  \begin{figure*}[tb]
    \centering
    \includegraphics[width=\textwidth]{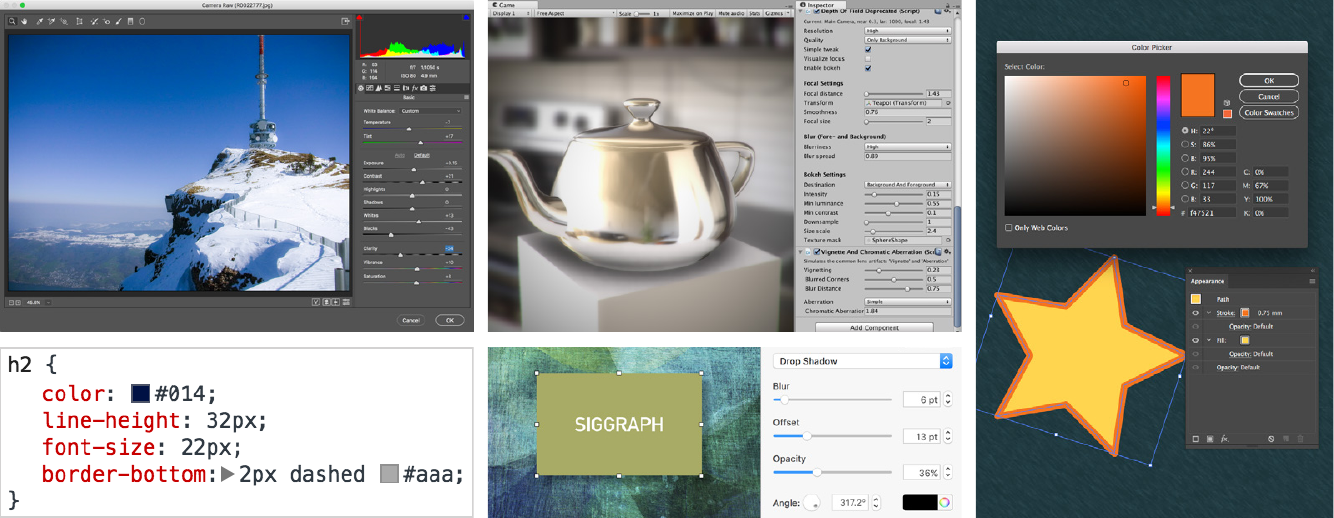}
    \caption[Example scenarios of parameter tweaking for visual design.]{
      \textbf{Example scenarios of parameter tweaking for visual design}, including photo color enhancement, image effects for 3D graphics, and 2D graphic designs, such as web pages, presentation slide.
    }
    \label{fig:sliders}
  \end{figure*}
}

\newcommand{\figgoodness}{
  \begin{figure*}[tb]
    \centering
    \includegraphics[width=\textwidth]{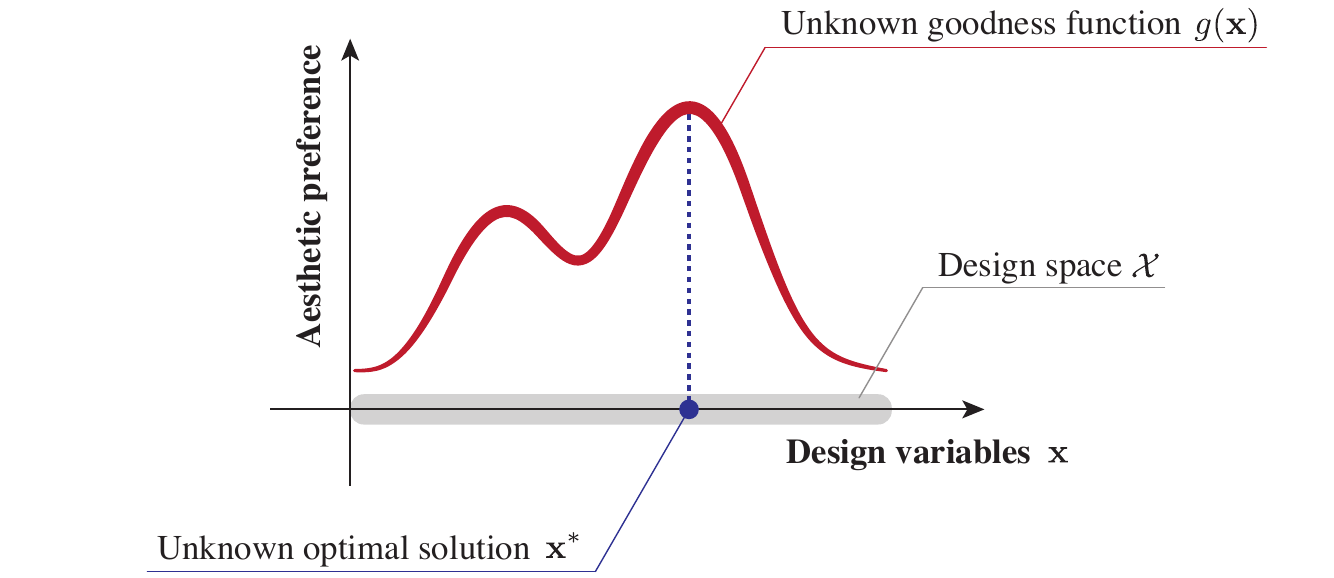}
    \caption[Problem formulation.]{
      \textbf{Problem formulation.}
      We discuss computational design methods to solve the optimization problem described in \autoref{eq:optimization} or to find the optimal solution $\bfx^{*}$ that maximizes the aesthetic preference of the target design.
    }
    \label{fig:goodness}
  \end{figure*}
}

\newcommand{\figtasconcept}{
  \begin{figure}[tb]
    \centering
    \includegraphics[width=\columnwidth]{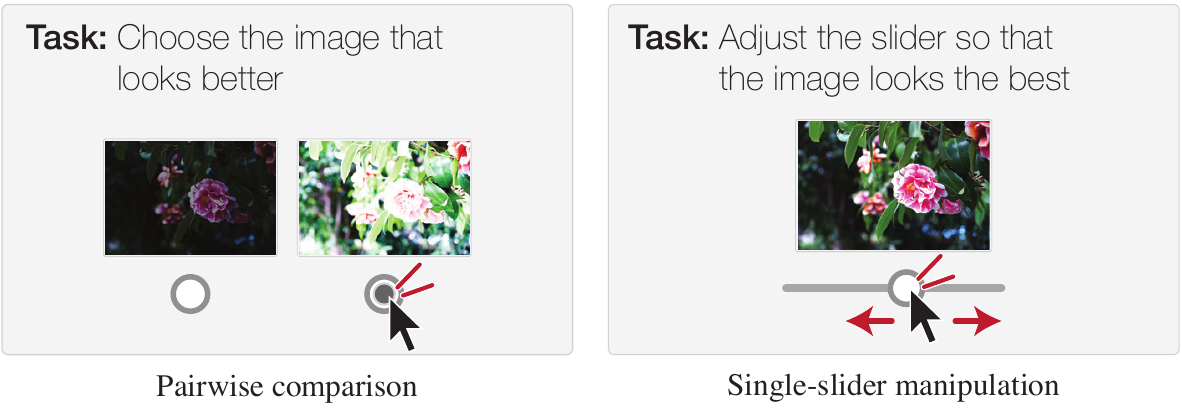}
    \caption[Microtask design.]{
      \textbf{Microtask design.}
      (Left) Pairwise comparison microtask.
      (Right) Single-slider manipulation microtask.
    }
    \label{fig:chapter5:tas_concept}
  \end{figure}
}

\newcommand{\figthreeteasersuggestion}{
  \begin{figure*}[tb]
    \centering
    \includegraphics[width=\textwidth]{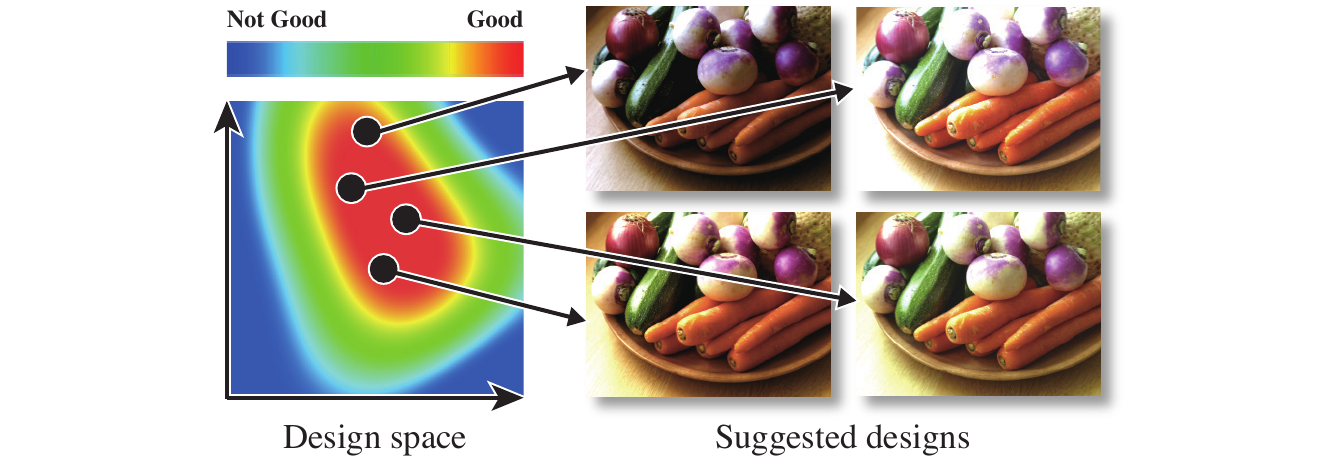}
    \caption[A suggestive interface enabled by the estimated goodness function.]{
      \textbf{A suggestive interface enabled by the estimated goodness function.}
      The user can obtain appropriate parameter sets as suggestions, which are generated considering the goodness of designs.
    }
    \label{fig:three:teaser:suggestion}
  \end{figure*}
}

\newcommand{\figthreeteaserslider}{
  \begin{figure*}[tb]
    \centering
    \includegraphics[width=\textwidth]{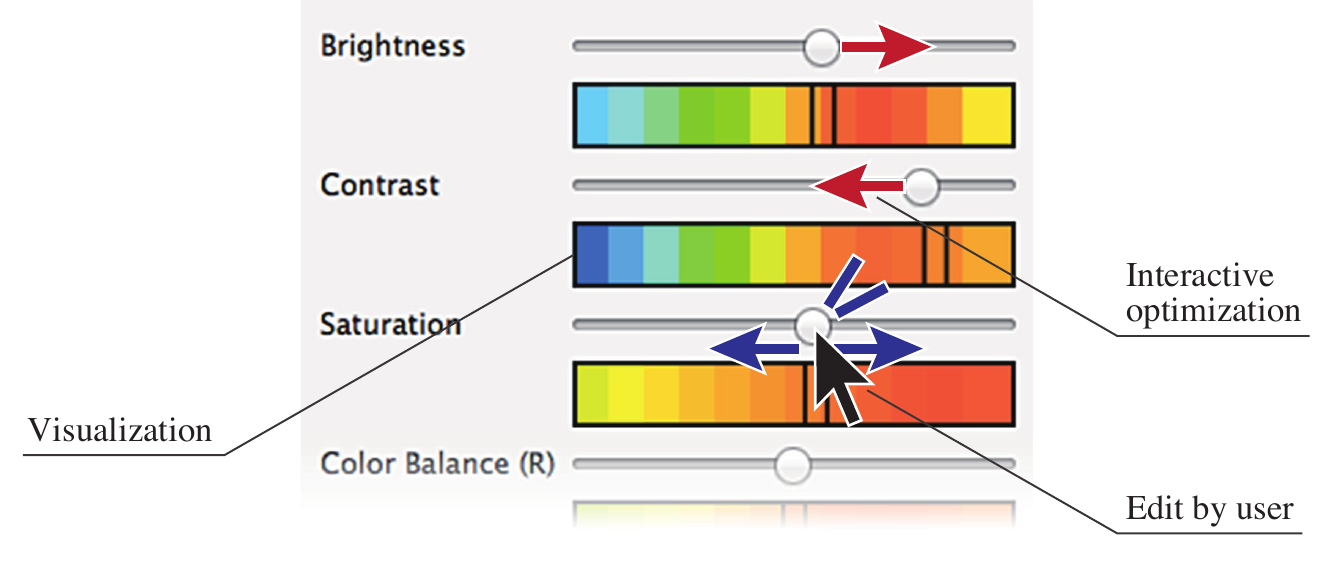}
    \caption[A slider interface enabled by the estimated goodness function.]{
      \textbf{A slider interface enabled by the estimated goodness function.}
      The user can adjust each parameter effectively by the visualization (Vis) near the slider and the optimization (Opt), which gently guides the current parameters toward the optimal direction.
    }
    \label{fig:three:teaser:slider}
  \end{figure*}
}

\newcommand{\figthreeoverview}{
  \begin{figure*}[tb]
    \centering
    \includegraphics[width=\textwidth]{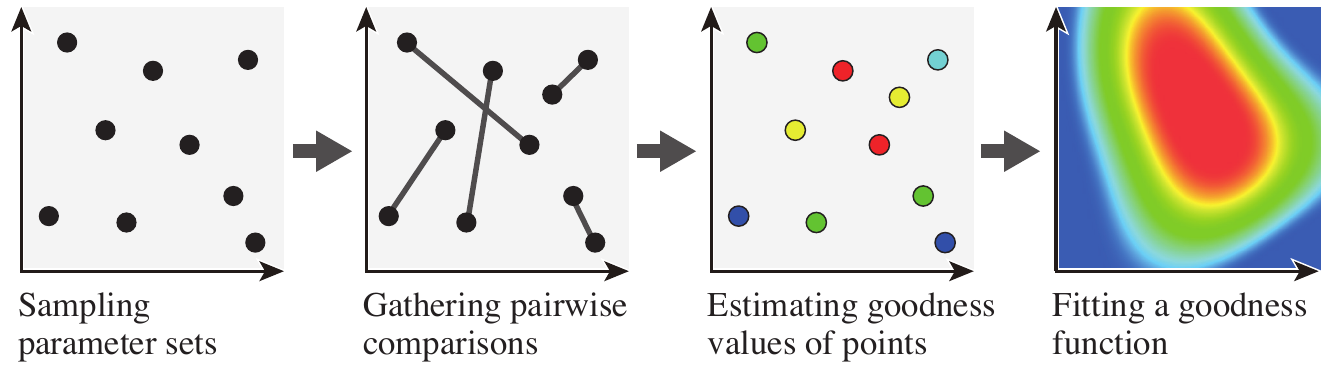}
    \caption[Overview of the crowd-powered algorithm for estimating the goodness function.]{
      \textbf{Overview of the crowd-powered algorithm for estimating the goodness function.}
    }
    \label{fig:three:overview}
  \end{figure*}
}

\newcommand{\figthreephoto}{
  \begin{figure}[tb]
    \centering
    \includegraphics[width=\linewidth]{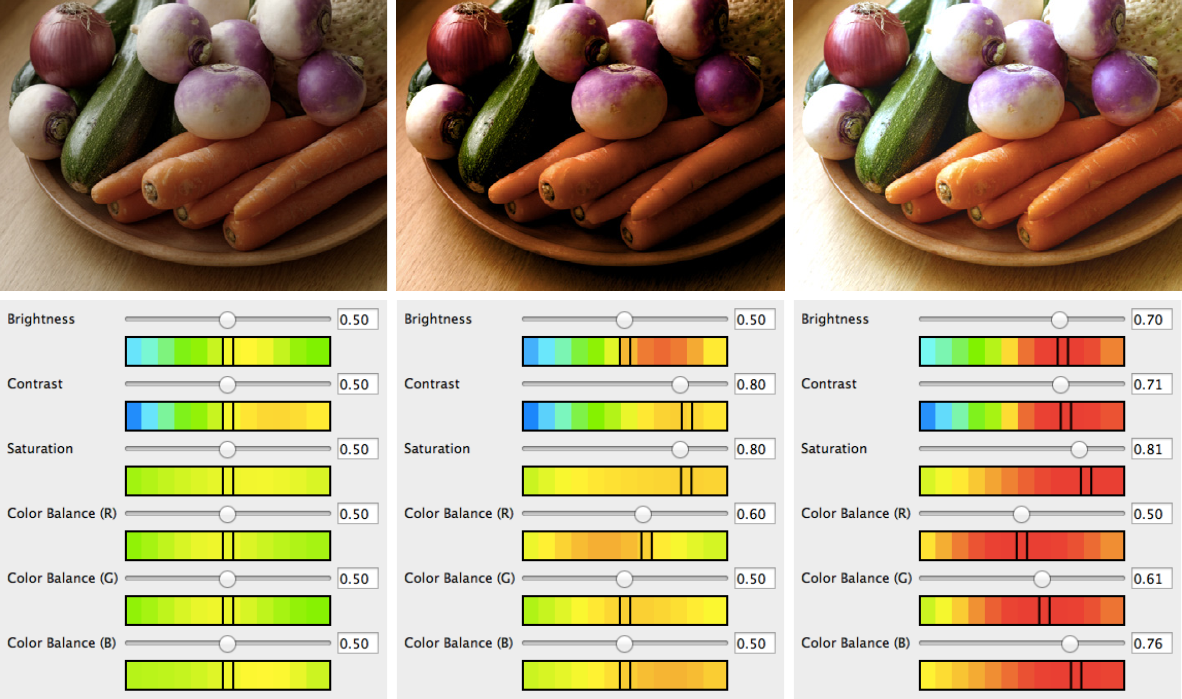}
    \caption[Designs and visualizations of goodness distributions in the photo color enhancement application.]{
      \textbf{Designs and visualizations of goodness distributions in the photo color enhancement application.}
    }
    \label{fig:three:photo}
  \end{figure}
}

\newcommand{\figthreeshader}{
  \begin{figure}[tb]
    \centering
    \includegraphics[width=\linewidth]{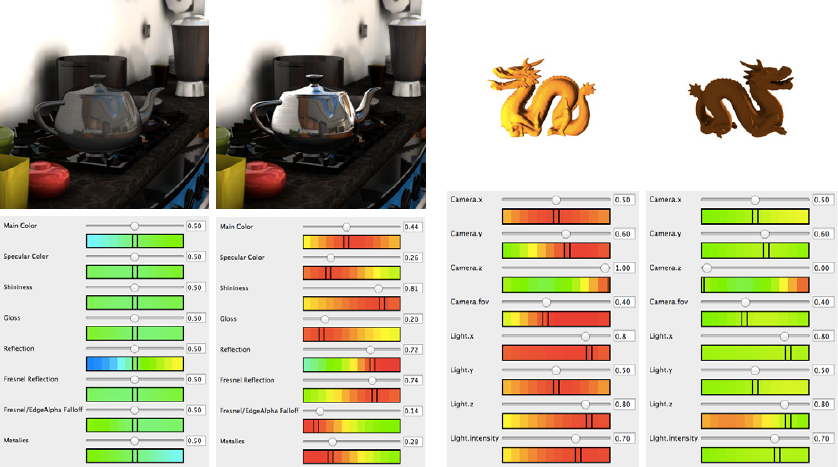}
    \caption[Designs and visualizations of goodness distributions in the shader application (Left) and the camera and light application (Right).]{
      \textbf{Designs and visualizations of goodness distributions in the shader application (Left) and the camera and light application (Right).}
    }
    \label{fig:three:shader}
  \end{figure}
}

\newcommand{\figthreeface}{
  \begin{figure*}[tb]
    \centering
    \includegraphics[width=\textwidth]{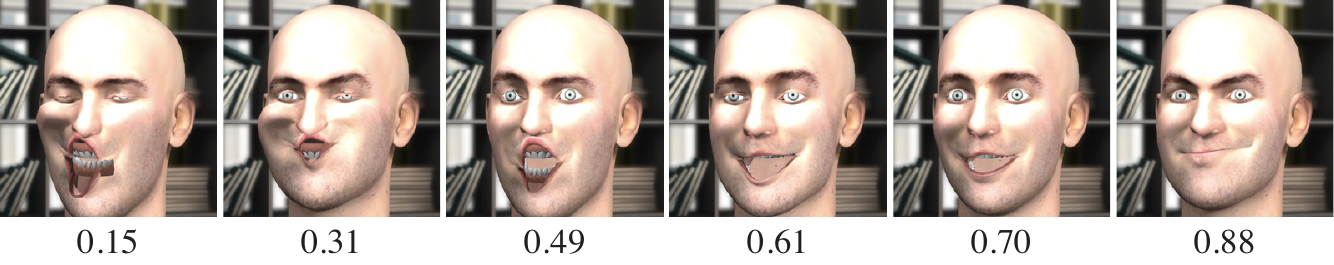}
    \caption[Designs and estimated goodness values in the facial expression modeling application.]{
      \textbf{Designs and estimated goodness values in the facial expression modeling application}.
      The values shown below the pictures are the estimated goodness values.
    }
    \label{fig:three:face}
  \end{figure*}
}

\newcommand{\figapproach}{
  \begin{figure*}[t]
    \includegraphics[width=\linewidth]{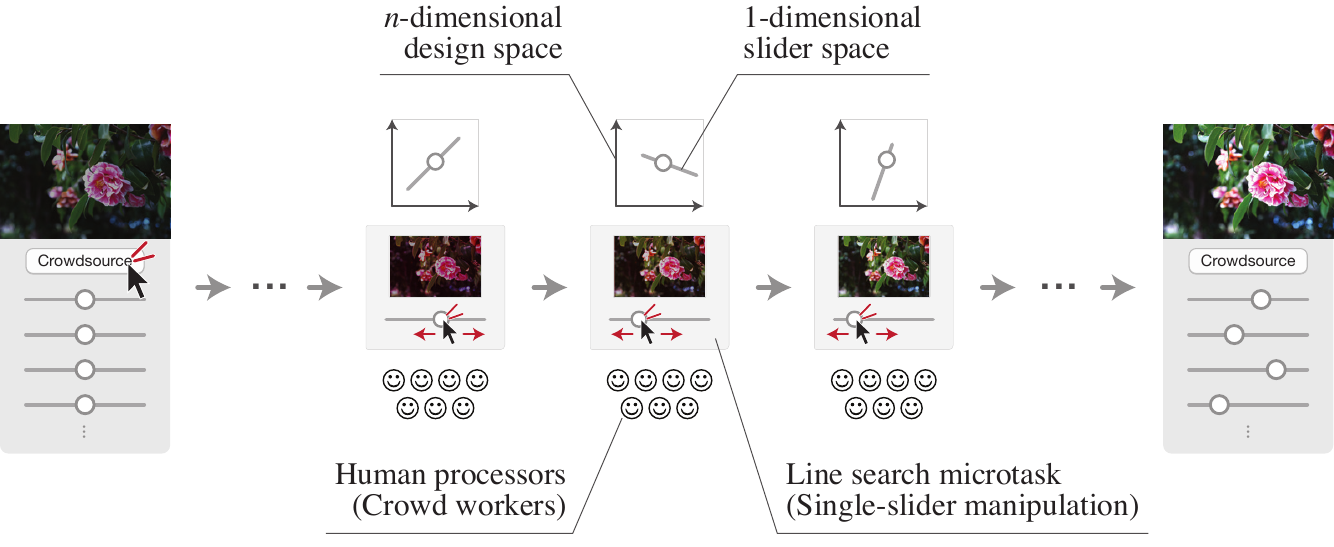}
    \caption[Concept of the method.]{
      \textbf{Concept of the method.}
      This work envisions that the design software equips a ``Crowdsource'' button for running the crowd-powered search for the slider values that provide the perceptually ``best'' design.
      To enable this, the system decomposes the $n$-dimensional optimization problem into a sequence of one-dimensional line search queries that can be solved by crowdsourced human processors.
    }
    \label{fig:chapter5:approach}
  \end{figure*}
}

\newcommand{\figsliderbayes}{
  \begin{figure}[tbp]
    \includegraphics[width=\columnwidth]{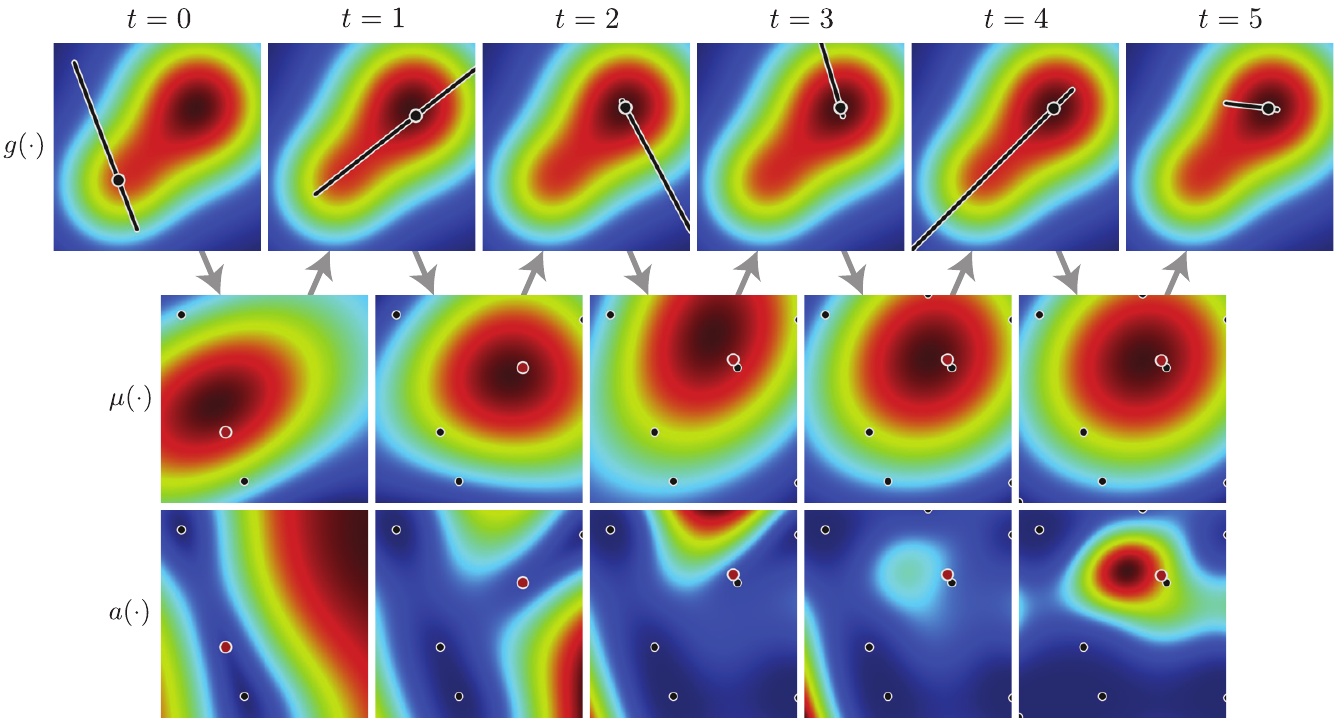}
    \caption[An example sequence of the Bayesian optimization based on line search oracle, applied to a two-dimensional test function.]{
      \textbf{An example sequence of the Bayesian optimization based on line search oracle, applied to a two-dimensional test function.}
      The iteration proceeds from left to right.
      From top to bottom, each row visualizes the black-box function $g(\cdot)$ along with the next slider space $\calS$ and the chosen parameter set $\bfx^\text{chosen}$, the predicted mean function $\mu(\cdot)$, and the acquisition function $a(\cdot)$, respectively.
      The red dots denote the best parameter sets $\bfx^{+}$ among the observed data points at each step.
    }
    \label{fig:chapter5:slider_bayes}
  \end{figure}
}

\newcommand{\figphotofree}{
  \begin{figure}[tb]
    \centering
    \includegraphics[width=\columnwidth]{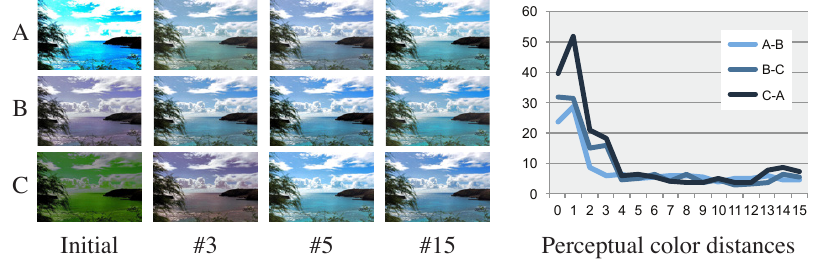}
    \caption[Comparison of three optimization trials with different initial conditions in photo color enhancement.]{
      \textbf{Comparison of three optimization trials with different initial conditions in photo color enhancement.}
      (Left) Transitions of the enhanced images.
      (Right) Transitions of the differences between each trial, measured by the perceptual color metric.
    }
    \label{fig:chapter5:photo_free}
  \end{figure}
}

\newcommand{\figphotophotoshop}{
  \begin{figure}[tb]
    \centering
    \includegraphics[width=\textwidth]{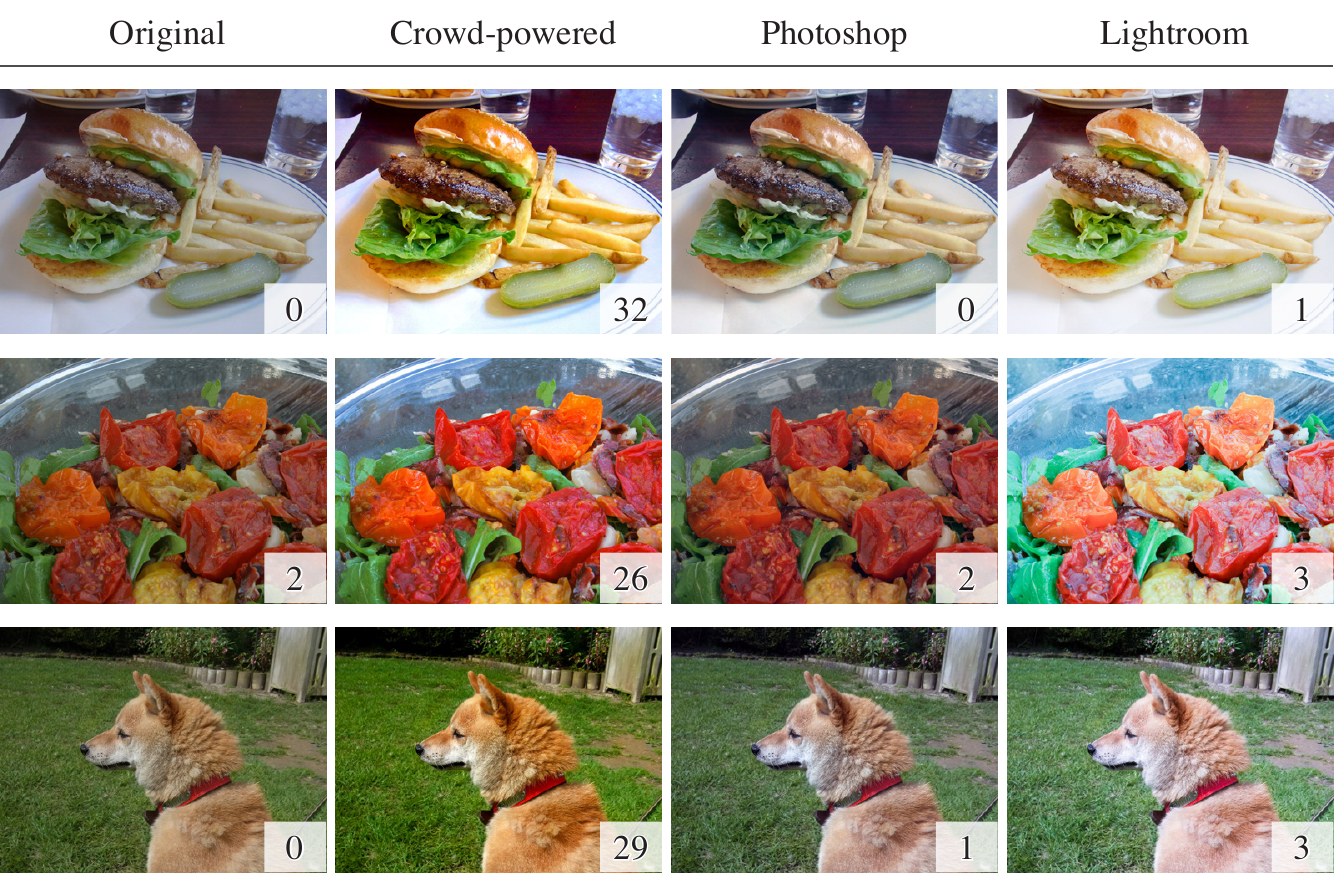}
    \caption[Comparison of photo color enhancement between our crowd-powered optimization and auto-enhancement in commercial software packages.]{
      \textbf{Comparison of photo color enhancement between our crowd-powered optimization and auto-enhancement in commercial software packages} (Photoshop and Lightroom).
      The number on each photograph indicates the number of participants who preferred the photograph to the other three in the study.
    }
    \label{fig:chapter5:photo_photoshop}
  \end{figure}
}

\newcommand{\figteapot}{
  \begin{figure*}[tbp]
  \centering
  \includegraphics[width=\textwidth]{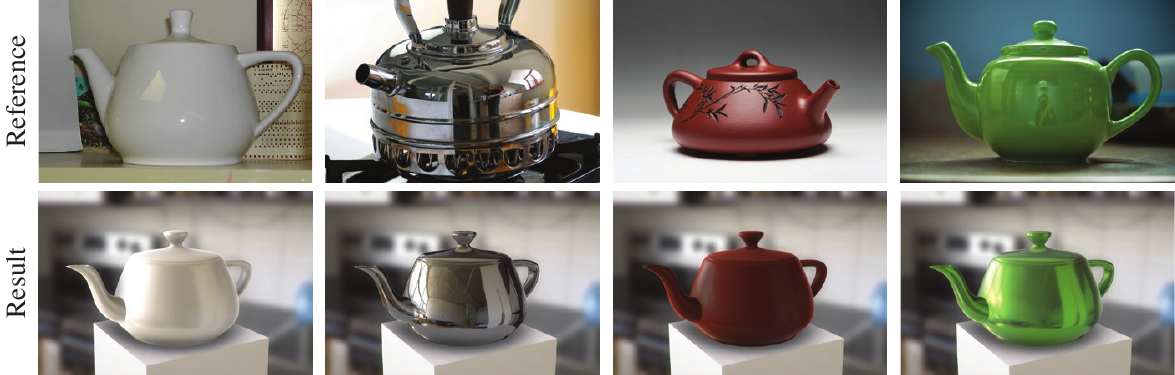}
  \caption[Results of the crowd-powered material appearance design with reference photographs.]{
    \textbf{Results of the crowd-powered material appearance design with reference photographs.}
    In each pair, the top image shows the reference photograph and the bottom image shows the resulting appearance.
    Some photographs were provided by Flickr users: Russell Trow, lastcun, and Gwen.
  }
  \label{fig:chapter5:teapot}
  \end{figure*}
}

\newcommand{\figteapotinstruction}{
  \begin{figure}
    \centering
    \includegraphics[width=0.8\columnwidth]{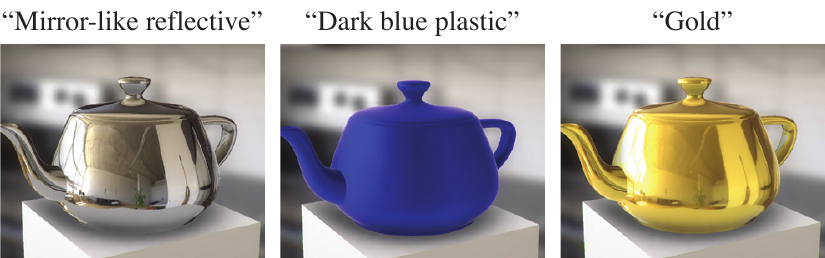}
    \caption[Results of the crowd-powered material appearance design with textual instructions.]{
      \textbf{Results of the crowd-powered material appearance design with textual instructions.}
    }
    \label{fig:chapter5:teapot_instruction}
  \end{figure}
}

\newcommand{\figsixcolors}{
  \begin{figure*}[tb]
    \centering
    \includegraphics[width=\textwidth]{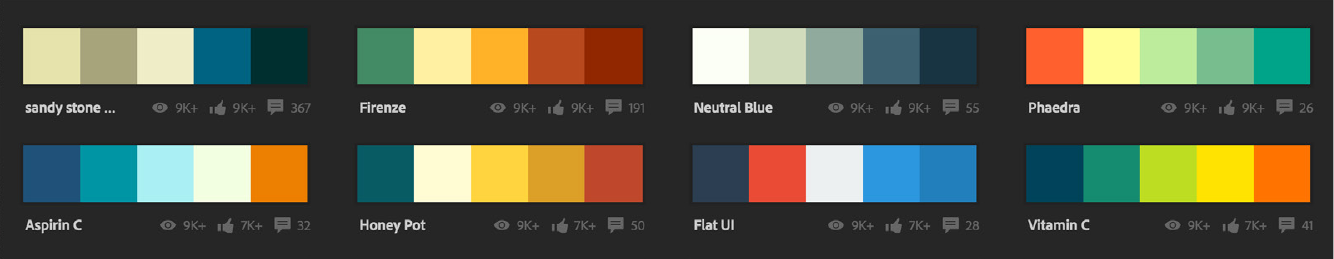}
    \caption[``Most Popular'' color palettes in the user community of Adobe Color CC.]{
      \textbf{``Most Popular'' color palettes in the user community of Adobe Color CC}.
      Though visually different from each other, they are (mostly) equally popular and preferred by many users.
    }
    \label{fig:chapter6:colors}
  \end{figure*}
}

\begin{document}

\maketitle

\begin{abstract}
  \setlength{\parskip}{0.6em}
  \setlength{\parindent}{0em}

  \noindent

Computational design is aimed at supporting or automating design processes using computational techniques.
However, some classes of design tasks involve criteria that are difficult to handle only with computers.
For example, visual design tasks seeking to fulfill aesthetic goals are difficult to handle purely with computers.
One promising approach is to leverage human computation; that is, to incorporate human input into the computation process.
Crowdsourcing platforms provide a convenient way to integrate such human computation into a working system.

In this chapter, we discuss such computational design with crowds in the domain of parameter tweaking tasks in visual design.
Parameter tweaking is often performed to maximize the aesthetic quality of designed objects.
Computational design powered by crowds can solve this maximization problem by leveraging human computation.
We discuss the opportunities and challenges of computational design with crowds with two illustrative examples:
(1) estimating the objective function (specifically, preference learning from crowds' pairwise comparisons) to facilitate interactive design exploration by a designer and
(2) directly searching for the optimal parameter setting that maximizes the objective function (specifically, crowds-in-the-loop Bayesian optimization).

  \hrulefill

  \begin{footnotesize}
    \renewcommand{\UrlFont}{\footnotesize\bf\ttfamily}

    \textbf{Citation.}
    Yuki Koyama and Takeo Igarashi. Computational design with crowds. In Antti Oulasvirta, Per Ola Kristensson, Xiaojun Bi, and Andrew Howes (Eds.), \textit{Computational Interaction}, chapter 6, pages 153--184. Oxford University Press, 2018. doi: \url{https://doi.org/10.1093/oso/9780198799603.001.0001}.

    \textbf{Definitive Version.}
    This material was originally published in Computational Interaction edited by Antti Oulasvirta, Per Ola Kristensson, Xiaojun Bi, and Andrew Howes, and has been reproduced by permission of Oxford University Press: \url{https://global.oup.com/academic/product/computational-interaction-9780198799603}. For permission to reuse this material, please visit \url{http://global.oup.com/academic/rights}.
  \end{footnotesize}
\end{abstract}

\tableofcontents


\section{Introduction}

\emph{Computational design} is the emerging form of design activities that are enabled by computational techniques.
It formulates design activities as mathematical optimization problems;
it formulates design criteria as either objective functions or constraints, design space as the search space (or the choice set), and design exploration, which can be performed by either systems, users, or their combination, as the process of searching for solutions.
This viewpoint provides an opportunity to devise new ways of utilizing computational techniques (\ie mathematical tools developed in computer science that leverage machine processing power).
The main goal of computational design research is to enable efficient design workflow or complex design outcomes that are impossible in traditional approaches relying purely on the human brain.

The quality of designed objects can be assessed using various criteria according to their usage contexts.
Some criteria might work as conditions that should be at least satisfied (\ie constraints);
other criteria might work as values that should be maximized (\ie objectives).
These design criteria can be classified into two groups: \emph{functional} criteria and \emph{aesthetic} criteria.
Functional criteria are the criteria about how well the designed object functions in the expected contexts.
For example, a chair is expected to be ``durable'' when someone is setting on it;
in this case, durability can be a functional criterion that should be satisfied.
In contrast, aesthetic criteria are about how perceptually preferable (or pleasing) the designed object looks.
A chair might look ``more beautiful,'' for example, if its shape is smooth and the width and height follow the golden ratio rule;
in this case, beauty in shape performs the role of an aesthetic criterion that is desired to be maximized.
Note that, in practical design scenarios, these two criteria are sometimes simultaneously considered by designers.

\subsection{Challenges in Aesthetic Design}

Design with aesthetic criteria is especially difficult for computers alone to handle.
The objective of a design problem is defined based on human perception, and it exists only in the brains of designers.
It is generally difficult to represent this perceptual objective using a simple equation or a rule that can be calculated by machine processors.
Thus, if one tries to devise a computational design method aiming at supporting or automating aesthetic design, it is inevitable for the framework to involve humans somewhere in the computation, which is non-trivial.
Another notable challenge is that humans, even designers, cannot consistently answer the goodness value for a certain design without being familiar with the design space (or the possible alternatives).
For example, suppose that you are shown a certain design and asked to provide its goodness score without knowing other possibilities;
the answer should be almost random.
However, if you are shown a certain design along with a baseline design, you could more reasonably provide its score by taking relative goodness into account.
This means that it is also non-trivial how to effectively query humans in the computation.

\subsection{A Solution: Crowd-Powered Methods}

A possible solution is to build computational design methods based on a \emph{human computation} paradigm, especially utilizing \emph{crowdsourcing} to involve \emph{crowd workers} in the design processes.
Crowdsourcing, especially in the form of \emph{microtask-based crowdsourcing}, enables computational systems to make use of the human perceptions of many people in a structured and algorithmic manner.
This idea allows researchers to devise new frameworks for supporting or even automating design processes that are otherwise difficult.
Crowdsourcing is superior to a single user or a small group in that the result is an average of many people, and thus it is less sensitive to individual variability and provides a reliable indicator of ``people's choice.''
It is also useful when the designer wants to design visuals optimized for a specific target audience with a cultural background different from that of the designer himself;
crowdsourcing platforms provide a convenient way for getting inputs from people with the target background.

The goals of this chapter are to discuss how to divide the task between human and computer, how to interact with crowds in the context of aesthetic design (\ie microtask design), how to extract mathematical information from the responses by crowds, how to utilize the information using computational techniques, and how to enhance design activities using computation (\ie interaction design).

This chapter specifically deals with \emph{parametric design} (\ie tweaking parameters for finding the best configuration) as a representative problem of aesthetic design.
For example, photo color enhancement (also referred to as tonal adjustment or color grading) is one of such design scenarios (\autoref{fig:aesthetics});
when a designer enhances the color of a photograph, he or she has to tweak multiple design parameters, such as ``brightness'' or ``contrast,'' via a slider interface to find the most aesthetically preferable enhancement for the target photograph.

In the following sections, we first provide the background and definitions of the concepts related to this chapter.
Then, we detail the discussion on computational design with crowds.
Next, we describe two illustrative examples of crowd-powered computational design methods.
Finally, we conclude this chapter with additional discussion on the remaining challenges and future directions.

\figaesthetics


\section{Background and Term Definitions}

This chapter provides discussions on \emph{computational design} methods for \emph{parametric design} problems by taking advantage of \emph{crowdsourced human computation}.
We first provide the definition of these terms and a short literature review.

\subsection{Computational Design}

We defined the term \emph{computational design} at the beginning of this chapter.
While we believe that our definition well explains this research field, we do not argue that ours is the sole definition.
As the term \emph{design} can be defined in many ways, computational design can also be defined in many ways.
Here, we review this field from the viewpoint of two design criteria: functional criteria and aesthetic criteria\footnote{Refer to Chapter 4 of this book \cite{Oulasvirta18} for a detailed discussion on design criteria for UI design.}.

\subsubsection{Functional Criteria}

Recently, many computational design methods for designing functional objects have been investigated, especially for digital fabrication applications.
So far, a variety of functional criteria have been formulated by researchers;
Umetani \etal \shortcite{Umetani14} formulated the functional criterion of paper airplanes (\ie \emph{fly-ability}) and used it for optimizing airplane designs by maximizing this criterion.
Koyama \etal \shortcite{Koyama15} formulated the \emph{hold-ability} and \emph{grip strength} of 3D-printed connectors and then presented an automatic method for designing functional connectors.
Several computational design methods consider the functionalities of objects' mass properties (\eg \emph{standing stability} \cite{Prevost13}).
\emph{Structural strength} of objects is also an important design criterion, and some computational design methods take this into consideration (\eg \cite{Stava12}).

Another notable domain of computational functional design is graphical user interface (GUI) generation.
In this context, the \emph{user performance} of the generated interface is often considered the functional criterion that should be optimized.
For example, Gajos \etal \shortcite{Gajos04} presented an automatic GUI design method, in which they formulated \emph{required user efforts} for manipulating GUI elements as the objective function to be minimized.
Bailly \etal \shortcite{Bailly13} presented MenuOptimizer, an interactive GUI design tool that utilizes an optimizer to support designers to design effective menus.

It is notable that these functional criteria are often ``computable'' by computers alone.
This is because these criteria are basically not subjective nor perceptual, in contrast to aesthetic criteria.

\subsubsection{Aesthetic Criteria}
\label{chapter:2:aesthetic_criteria}

The goal of computational design for aesthetic criteria is to support or automate the maximization of the perceptual aesthetic quality of designs.
Such aesthetic preference is closely tied to human perception, and thus it is difficult to quantify using simple rules.
Yet, by focusing on very specific design domains, it is possible to handle and optimize aesthetic criteria by rule-based approaches.
For example, Miniukovich and Angeli \shortcite{Miniukovich15} presented executable metrics of GUI aesthetics and showed its validity with a user study.
Todi \etal \shortcite{Todi16} applied similar metrics for optimizing GUI layout design.
However, rule-based approaches require careful implementation and tuning of heuristic rules in limited scopes.

Data-driven approaches can ease the limitations of purely rule-based approaches.
Most data-driven methods rely on heuristic rules but can derive optimal weights or model parameters for the rules by learning them from training data.
For example, O'Donovan \etal \shortcite{ODonovan14TVCG} presented a data-driven method of predicting and optimizing aesthetic quality of layouts of two-dimensional graphic designs.
Their aesthetic criterion is formulated by combining several heuristic rules (\eg \emph{alignment} and \emph{white space}), and machine learning techniques are used to learn the weights and the model parameters.
Other examples deal with color palette aesthetics \cite{ODonovan11}, 3D viewpoint preference \cite{Secord11}, and photo color enhancement \cite{Bychkovsky11}.

Talton \etal \cite{Talton09} presented a method for supporting preference-driven parametric design by involving many people.
Their method constructs a so-called \emph{collaborative design space}, which is a subset of the target design parameter space consisting of aesthetically acceptable designs, based on the design history of many voluntary users.
Then, the collaborative design space supports new users' design exploration.
Their method takes roughly one year to obtain the necessary design history and needs many volunteers to engage exactly the same design space.
In contrast, more recent crowdsourcing platforms have enabled \emph{on-demand} generation of necessary data, which has opened new opportunities for computational aesthetic design.

\subsection{Crowdsourcing and Human Computation}

Human computation and crowdsourcing are often used for gathering human-generated data that are difficult for machines to generate (\eg perceptual or semantic labels for images).
We utilize this approach for formulating our crowd-powered methods for gathering perceptual preference data.
We encourage readers to refer to the comprehensive survey and discussions on these terms by Quinn and Bederson \shortcite{Quinn11}.
Here, we briefly review these two terms from the viewpoint of our attempt.

\subsubsection{Human Computation}

Human computation is a concept of enabling difficult computations by exploiting humans as processing powers.
This term was described by von Ahn \shortcite{VonAhn05} as follows:
\begin{quote}
\ldots a paradigm for utilizing human processing power to solve problems that computers cannot yet solve.
\end{quote}
For example, human processors are much better at perceiving the semantic meanings of visual contents than machine processors;
thus, for building a system that requires perceptive abilities, it is effective to incorporate human processors as well as machine processors.
Such problems that are difficult for machine processors but easy for human processors, including visual design driven by preference, are observed in many situations.
However, human processors also have critical limitations, such as that they are extremely slow and expensive to execute compared to machine processors.
Therefore, it is important to carefully choose how to employ such human processors.

\paragraph{How to Employ Human Processors.}

A possible solution for employing many human processors is to implicitly embed human computation tasks in already existing tasks.
reCAPTCHA \cite{vonAhn08Science} takes such an approach; it embeds optical character recognition (OCR) tasks to web security measures.
Unfortunately, this approach is not easy for ordinary people to take.
Another solution to motivate many ordinary people to voluntarily participate in human computation tasks is to do ``gamification'' of tasks so that people do tasks purely for entertainment purpose \cite{vonAhn08ACMComm}.
For example, ESP game \cite{vonAhn04} is a game in which players provide semantic labels for images without being aware of it.
Recently, since the emergence of large-scale crowdsourcing markets, it has become increasingly popular to employ human processors using crowdsourcing.
As we take this approach, we detail it in the following subsections.

\subsubsection{Crowdsourcing}

The term \emph{crowdsourcing} was first introduced by Howe \cite{Howe06} in 2006 and later more explicitly defined in \cite{Howe06b} as follows.
\begin{quote}
Crowdsourcing is the act of taking a job traditionally performed by a designated agent (usually an employee) and outsourcing it to an undefined, generally large group of people in the form of an open call.
\end{quote}
Today, many online marketplaces for crowdsourcing are available for researchers, such as Upwork\footnote{\url{https://www.upwork.com/}}, Amazon Mechanical Turk\footnote{\url{https://www.mturk.com/}}, and CrowdFlower\footnote{\url{https://www.crowdflower.com/}}.
Since 2006, crowdsourcing has been a more and more popular research topic in computer science.
The forms of crowdsourcing are roughly categorized into the following categories.
\begin{description}
\item[Microtask-Based Crowdsourcing] is a form of crowdsourcing where many workers are employed to perform a \emph{microtask}---a task that is very small (usually completed in a minute) and does not require any special skills or domain knowledge to be performed.
One of the most attractive features of this form is that anyone can stably employ a large number of crowd workers even for small tasks on demand without any communication cost.
Recent crowdsourcing marketplaces, including Amazon Mechanical Turk, have enabled this new form.
Although crowd workers in these platforms are usually non-experts, they do have full human intelligence, which enables many emerging applications.

One of the popular usages of microtask-based crowdsourcing is to outsource data-annotation tasks (\eg \cite{Bell13}) for machine learning purposes.
Another popular usage is to conduct large-scale perceptual user studies (\eg \cite{Kittur08}).
Microtask-based crowdsourcing also enables \emph{crowd-powered systems}, which are systems that query crowd workers to use their human intelligence in run time.
For example, Soylent \cite{Bernstein10} is a crowd-powered word processing system that utilizes human intelligence to edit text documents.
\item[Expert Sourcing] is a form of crowdsourcing in which skilled experts (\eg web developers, designers, and writers) are employed for professional tasks.
Some online marketplaces (\eg Upwork) provide an opportunity for researchers to reach experts.
However, asking professional designers takes significant communication costs and large variances between individuals' skills.
An expert with the required skill is difficult to find and not always available.
Compared to microtask-based crowdsourcing, expert sourcing may be less suitable for employing human processors and for designing crowd-powered systems that work stably and on demand.
\item[Volunteer-Based Crowdsourcing] is a form of crowdsourcing in which unpaid crowds voluntarily participate in the microtask execution \cite{Huber17,Morishima14}.
One challenge in this form is to motivate crowd workers;
crowds would do not tend to perform tasks unless the task execution provides certain values other than monetary rewards (\eg public recognition for the efforts).
\end{description}

Researchers on crowdsourcing have investigated many issues.
For example, Bernstein \etal \shortcite{Bernstein11} proposed a technique for reducing the latency of responses from crowds for enabling real-time crowd-powered systems.
\emph{Quality control} of workers' responses \cite{Ipeirotis10} is also an important issue because crowd workers might make poor-quality responses because of cheating, misunderstanding of tasks, or simply making mistakes.
In this chapter, we do not discuss these issues and instead focus on the mechanism of how crowds can contribute to design activities.

\subsubsection{Crowdsourced Human Computation}
\label{chapter:2:crowdsourced_human_computation}

We define \emph{crowdsourced human computation} as a form of human computation in which human processors are employed via microtask-based crowdsourcing.
This means that programmers can query ``oracles'' requiring human intelligence into their codes as function calls.
Little \etal \shortcite{Little10} presented \emph{TurKit Script}, a programming API for developing algorithms using crowdsourced human computation (which they call \emph{human computation algorithms}).

Gingold \etal \shortcite{Gingold12} proposed several methods for solving long-standing visual perceptual problems using crowdsourced human computation, including the extraction of depth and normal maps for images and the detection of bilateral symmetries in photographs.
Their image-understanding algorithms are designed to \emph{decompose} the original difficult problem into a set of easy perceptual microtasks, \emph{solve} the perceptual microtasks using crowdsourcing, and then \emph{recompose} the responses from crowds using some computational techniques.

\subsection{Parametric Design}
\label{chapter:2:parametric}

In this chapter, we take \emph{parametric design} as a representative task in aesthetic design.
We use the term \emph{parametric design} to represent a design paradigm in which visual contents are solely controlled by a set of (either continuous or discrete) parameters.
Also, in parametric design, the number of parameters is often reasonably small so that designers can manually tweak them.
For example, to adjust the tone of a photograph, designers tweak several sliders, such as ``brightness'' and ``contrast,'' rather than tweaking the RGB values of every pixel one by one;
it is considered that the design space here is parametrized by several degrees of freedom mapped to sliders, and thus it is considered a parametric design.

Parametric designs can be found almost everywhere in visual design production.
\autoref{fig:sliders} illustrates a few examples in which the visual content is tweaked so that it becomes aesthetically the best.
For example, in Unity\footnote{\url{https://unity3d.com/}} (a computer game authoring tool) and Maya\footnote{\url{https://www.autodesk.com/products/maya/overview}} (a three-dimensional computer animation authoring tool), the control panels include many sliders, which can be manipulated to adjust the visual nature of the contents.

Finding the best parameter combination is not an easy task.
It may be easily found by a few mouse drags in a case in which the designer is familiar with how each parameter affects the visual content and is very good at predicting the effects without actually manipulating sliders.
However, this is unrealistic in most cases;
several sliders mutually affect the resulting visuals in complex ways, and each slider also has a different affect when the contents are different, which make the prediction difficult.
Thus, in practice, it is inevitable that a designer explores the \emph{design space}---the set of all the possible design alternatives---in a trial-and-error manner, to find the parameter set that he or she believes are best for the target content.
This requires the designer to manipulate sliders many times, as well as to construct a mental model of the design space.
Furthermore, as the number of design parameters increases, the design space expands exponentially, which makes this exploration very tedious.

\figsliders

Computer graphics researchers have investigated many methods for defining reasonable parametric design spaces.
For 3D modeling, the human face \cite{Blanz99} and body \cite{Allen03} are parametrized using data-driven approaches.
The facial expression of characters is often parametrized using \emph{blendshape} techniques \cite{Lewis14}.
For material appearance design, Matusik \etal \shortcite{Matusik03} proposed a parametric space based on measured data, and Nielsen \etal \shortcite{Nielsen15} applied dimensionality reduction to the space.
Procedural modeling of 3D shapes (\eg botany \cite{Weber95}) is also considered parametric design in that the shapes are determined by a set of tweakable parameters.
One of the recent trends is to define parametric spaces based on semantic attributes for facilitating intuitive exploration;
this direction has been investigated for shape deformation \cite{Yumer15TOG}, cloth simulation \cite{Sigal15}, and human body shape \cite{Streuber16}, for example.


\section{Computational Design with Crowds}
\label{sec:concept}

\subsection{Problem Formulation}

We consider a design exploration in which multiple design parameters have to be tweaked such that the aesthetic preference criterion is maximized.
This can be mathematically described as follows.
Suppose that there are $n$ real-valued design variables
\begin{equation}
\bfx = \begin{bmatrix} x_1 & \cdots & x_n \end{bmatrix} \in \calX,
\end{equation}
where $\calX$ represents an $n$-dimensional design space.
We assume that
\begin{equation}
\calX = [0, 1]^n;
\end{equation}
that is, each variable takes a continuous value and its interval is regularized into $[0, 1]$ in advance.
In the case that a variable is manipulated by a slider, we suppose that the slider's lowest and highest values correspond to $0$ and $1$, respectively.
Using these notations, a parameter-tweaking task can be formulated as a continuous optimization problem:
\begin{equation}
\bfx^{*} = \amax{\bfx \in \calX} g(\bfx),
\label{eq:optimization}
\end{equation}
where the objective function
\begin{equation}
g : \calX \rightarrow \realnumber
\end{equation}
returns a scalar value representing how aesthetically preferable the design corresponding to the argument design variables is.
We call this function as \emph{goodness function}.
A designer usually tries to solve this optimization problem by exploring $\calX$ by manipulating sliders in a trial-and-error manner without any computational support.
This exploration ends when the designer believes that he or she finds the optimal argument value $\bfx^{*}$ that gives the best preferable design.
\autoref{fig:goodness} illustrates this problem setting.
The goal of this chapter is to provide discussions for seeking methods for solving this optimization problem either automatically or semi-automatically using computational techniques.

\figgoodness

\paragraph{Assumptions and Scope.}

We focus our discussions on parametric design and put several assumptions on this problem setting to clarify the scope in this discussion.
First, we assume that the target parameters are continuous, not discrete.
We also assume that when a design parameter changes smoothly, the corresponding visual also changes smoothly.
From these assumptions, the goodness function $g(\cdot)$ is considered to be a continuous, smooth function.
Note that the goodness function can have multiple local maximums, ridges, or locally flat regions around maximums.
Also, we assume that the goodness function is constant with respect to time.
The design space is expected to be parameterized by a reasonable number of parameters as in most commercial software packages;
parametrization itself will be not discussed in this chapter.
We handle the design domains in which even novices can assess relative goodness of designs (for example, given two designs, they are expected to be able to answer which design looks better);
but importantly, they do not need to know how a design can be improved.
Though we narrow down the target problem as discussed above, it still covers a wide range of practical design scenarios including photo enhancement, material appearance design for computer graphics, and two-dimensional graphic design (\eg posters).

\subsection{Where to Use Computational Techniques}

To provide computational methods for solving the design problem described as \autoref{eq:optimization}, there are two possible approaches with respect to the usage of computational tools as follows.
\begin{description}
\item[Estimation of the Objective Function.]
The first approach is to estimate the shape of the goodness function $g(\cdot)$ by using computational tools.
In other words, it is to compute the regression of $g(\cdot)$.
Once $g(\cdot)$ is estimated, it can be used for ``guided'' exploration: supporting users' free exploration of the design space $\calX$ for finding their best favorite parameter set $\bfx^{*}$ through some user interfaces.
One of the advantages of this approach is that even if the estimation quality is not perfect, it can still be effective for supporting users to find $\bfx^{*}$.
To implement this approach, there are two important challenges:
how to compute this regression problem that deals with human preference and how to support the users' manual exploration using the estimated $g(\cdot)$.
We discuss this approach further in \autoref{sec:uist}.
\item[Maximization of the Objective Function.]
The second approach is to compute the maximization of the goodness function $g(\cdot)$ by using computational optimization techniques so that the system can directly find the optimal solution $\bfx^{*}$.
In other words, the system searches the design space $\calX$ for the maximum of $g(\cdot)$.
The found solution $\bfx^{*}$ can be used as either a final design or a starting point that will be further refined by the user.
Implementing this approach requires several non-trivial considerations, for example, which optimization algorithms can be used and how it should be adapted for this specific problem setting.
We discuss this approach further in \autoref{sec:siggraph}.
\end{description}

For both approaches, human-generated preference data are necessary for enabling computation.
In this chapter, we focus on the use of human computation to generate necessary preference data.
Such human processors can be employed via crowdsourcing.
By this approach, systems can obtain crowd-generated data on demand in the manner of function calls.
Here we put an additional assumption: a common ``general'' preference exists that is shared among crowds;
though there might be small individual variation, we can observe such a general preference by involving many crowds.

\subsection{What and How to Ask Crowds}
\label{sec:microtask}

It is important to ask appropriate questions to the crowds to obtain meaningful results.
We discuss two possibilities in terms of the query design as follows.
\begin{description}
\item[Query about Discrete Samples.]
A typical approach is to ask crowds about the aesthetic quality of some discrete samples.
The simplest task design is to present a sample in the parameter space and ask crowd workers to answer its goodness score.
An interface for crowd workers could be a slider, a text box (for putting the score value), or an $n$-pt Likert questionnaire.
From a mathematical viewpoint, this can be modeled as follows:
given a parameter set $\bfx$ that the system is inspecting, crowds provide the function value
\begin{equation}
g(\bfx).
\end{equation}
However, this does not work well in general.
It requires crowd workers to be familiar with the design space;
otherwise, it is difficult for crowd workers to make answers consistently (as also discussed at the beginning of this chapter).

A solution is to present multiple samples to crowds and to ask them to evaluate their ``relative'' goodness.
For example, it is easy for crowds to choose the better design from two options (\ie a pairwise comparison).
The interface for this task could be a simple radio button.
\autoref{fig:chapter5:tas_concept} (Left) illustrates this microtask design.
This can be mathematically modeled as follows:
given two sampling points, say $\bfx_1$ and $\bfx_2$, crowds provide the information of their relative order
\begin{equation}
g(\bfx_1) < g(\bfx_2),
\end{equation}
or its opposite.
This task is easy to answer even for non-experts because it does not require the crowds to know the design space or other design possibilities.
A possible variant of this task is to provide relative scores by an $n$-pt Likert scale (the standard pairwise comparison is the special case with $n = 2$).
A/B testing also belongs to this category.
This pairwise comparison microtask is popular in integrating crowds' perceptions into systems (\eg \cite{ODonovan14TOG,Gingold12}).
We discuss this approach further in \autoref{sec:uist}.

\figtasconcept

\item[Query about Continuous Space.]
An alternative approach is to ask crowd workers to explore a continuous space and to identify the best sample in the space.
This requires more work by crowds, but the system can obtain much richer information than an evaluation of discrete samples.
Asking crowds to directly control all the raw parameters (\ie explore the original search space $\calX$) is an extreme case of this approach.
Mathematically speaking, given a high-dimensional parametric space $\calX$, crowds provide the solution of the following maximization problem:
\begin{equation}
\amax{\bfx \in \calX} g(\bfx).
\end{equation}
However, an exploration of such a high-dimensional space is highly difficult for crowd workers.
Note that this task is difficult even for designers.
Also, as this task is no longer ``micro,'' it is less easy to stably obtain reliable quality responses.

A solution is to limit the search space to a lower-dimensional space.
For example, by limiting the space into a one-dimensional subspace, crowd workers can efficiently explore the space with a single slider.
\autoref{fig:chapter5:tas_concept} (Right) illustrates this microtask design.
This is mathematically formulated as the following maximization problem:
\begin{equation}
\amax{\bfx \in \calS} g(\bfx),
\end{equation}
where $\calS$ is a (one-dimensional) continuous space that can be mapped to a single slider.
We discuss this approach further in \autoref{sec:siggraph}.
\end{description}

These microtask designs and their mathematical interpretations enable computational systems to access human preferences in the algorithms and to incorporate them into the mathematical formulations.
In the following sections, we introduce two such crowd-powered systems from our previous studies as illustrative examples.


\section{Design Exploration with Crowds}
\label{sec:uist}

In this section, we illustrate a crowd-powered method to facilitate parametric design exploration \cite{Koyama14}.
To gather the necessary preference data, the system asks crowds to perform pairwise comparison microtasks.
The system then analyzes the data to estimate the landscape of the goodness function.
The estimated goodness function is used to enable novel user interfaces for facilitating design exploration.

\subsection{Interaction Design}

The system provides the following two user interfaces to support design exploration leveraging the goodness function obtained by inputs from crowds.
\begin{description}
\item[Suggestive Interface.]
The system has a suggestive interface called Smart Suggestion (\autoref{fig:three:teaser:suggestion}).
It generates nine parameter sets that have relatively high goodness values and displays the corresponding designs as suggestions.
This interface facilitates design exploration by giving users a good starting point to find a better parameter set for the visual design.
This implementation takes a simple approach to generate quality suggestions:
the system generates 2,000 parameter sets randomly and then selects the nine best parameter sets according to their goodness values.
This simple algorithm interactively provides suggestions of an adequate quality, which enables users to regenerate suggestions quickly enough for interactive use if none of the suggestions satisfy them.
\item[Slider Interface.]
The system has a slider interface called VisOpt Slider (\autoref{fig:three:teaser:slider}).
It displays colored bars with a \emph{visualization} of the results of a crowd-powered analysis.
The distribution of goodness values is directly visualized on each slider using color mapping, which navigates the user to explore the design space.
Note that when the user modifies a certain parameter, the visualizations of the other parameters change dynamically.
This helps the user not only to find better parameter sets quickly but also to explore the design space effectively without unnecessarily visiting ``bad'' designs.
When the \emph{optimization} is turned on, the parameters are automatically and interactively optimized while the user is dragging a slider.
That is, when a user starts to drag a slider, the other sliders' ticks also start to move simultaneously to a better direction according to the user's manipulation.
\end{description}
These two interfaces are complementary;
for example, a user can first obtain a reasonable starting point by the suggestive interface and then interactively tune it up using the slider interface.

\figthreeteasersuggestion
\figthreeteaserslider

\subsection{Estimating the Objective Function}

The system employs crowdsourcing to analyze parameters to support the exploration of visual design.
The goal of the analysis is to construct the goodness function $g(\cdot)$.
The process to obtain a goodness function consists of four steps, as shown in \autoref{fig:three:overview}.

\figthreeoverview

\paragraph{Sampling Parameter Sets.}

First, the system samples $M$ parameter sets $\bfx_1,\ldots,\bfx_M$ from the parameter space $\calX$ for the later process of crowdsourcing.
To do this, we simply choose a random uniform sampling;
the system randomly picks a parameter set from the parameter space and repeats this process $M$ times.

\paragraph{Gathering Pairwise Comparisons.}

The next step is to gather information on the goodness of each sampling point.
We take the pairwise comparison approach; crowd workers are shown a pair of designs and asked to compare them.
As a result, relative scores (instead of absolute ones) are obtained.
For the instruction of the microtask for crowd workers, we prepare a template:
\begin{quote}
\textit{
Which of the two images of \emph{[noun]} is more \emph{[adjective]}? For example, \emph{[clause]}. Please choose the most appropriate one from the 5 options below.
}
\end{quote}
In accordance with the purpose and the content, the user gives a noun, an adjective such as ``good'' or ``natural,'' and a clause that explains a concrete scenario to instruct crowd workers more effectively.
After this instruction, two images and five options appear.
These options are linked to the 5-pt.\ Likert scale;
for example, ``\textit{the left image is definitely more \emph{[adjective]} than the right image}'' is for option 1, and the complete opposite is option 5.
Option 3 is ``\textit{these two images are equally \emph{[adjective]}, or are equally not \emph{[adjective]}.}''

\paragraph{Estimating Goodness Values of Sampling Points.}

Given the relative scores, the next goal is to obtain the absolute goodness values
$\bfy = [ \, y_{1} \:\: \cdots \:\: y_{M} \, ]^T$
at the sampling points $\bfx_{1}, \ldots, \bfx_{{M}}$.
Note that inconsistency exists in the data from crowds;
thus, a solution that satisfies all the relative orders does not generally exist.
We want to obtain a solution that is as reasonable as possible.
In this work, this problem is solved by being formulated as optimization.
Specifically, the system solves the optimization:
\begin{equation}
\min_{\bfy} \left\{ E_{\textrm{relative}} (\bfy) + \omega E_{\textrm{continuous}} (\bfy) \right\},
\label{minimization}
\end{equation}
where $E_\textrm{relative}(\cdot)$ is a cost function that reflects the relative scores data provided by crowds, $E_\textrm{continuous}(\cdot)$ is a cost function that regularizes the resultant values so as to be distributed smoothly and continuously, and $\omega > 0$ is a hyperparameter that defines the balance of these two objectives.

\paragraph{Fitting a Goodness Function.}

Now, the goal is to obtain a continuous goodness function from the goodness values at the discrete sampling points obtained in the previous step.
For this purpose, this work adopted the radial basis function (RBF) interpolation technique \cite{Bishop95}, which can be used to smoothly interpolate the values at scattered data points.

\subsection{Applications}

This method was applied to four applications from different design domains.
In this experiment, each microtask contains 10 unique pairwise comparisons, and 0.02 USD is paid for it.
For the photo color enhancement (6D), material appearance design (8D), and camera and light control (8D) applications, we deployed 200 tasks.
For the facial expression modeling (53D) application, we deployed 600 tasks.

\subsubsection{Photo Color Enhancement}

Here we selected six popular parameters for photo color enhancement:
brightness, contrast, saturation, and color balance (red, green, and blue).
In the crowdsourced microtasks, we asked crowd workers to choose the photograph that would be better to use in a magazine or product advertisement.
Examples of VisOpt Slider visualizations with typical parameter sets are shown in \autoref{fig:three:photo}.
These visualizations provide assorted useful information;
for example, the photo at left needs to have higher contrast, the center photo can be improved by making the brightness slightly higher and the red balance slightly lower, and the right photo is already good and does not require any dramatic improvements.

\figthreephoto

\subsubsection{Material Appearance Design}

Material appearance design is often difficult for novices to understand and tweak.
The problem is that shaders often have unintuitive parameters that affect the final look in a way that is difficult for casual users to predict (\eg ``Fresnel Reflection'').
For this experiment, we used a shader for photo-realistic metals provided in a popular shader package called Hard Surface Shaders\footnote{\url{https://www.assetstore.unity3d.com/\#/content/729}}, which has eight parameters.
We applied this shader to a teapot model.
We asked crowd workers to choose the one that was the most realistic as a stainless steel teapot.
\autoref{fig:three:shader} (Left) shows typical parameter sets with their visualizations.
From these visualizations, we can learn, without any trial and error, that the ``Reflection'' parameter (the fifth parameter in \autoref{fig:three:shader} (Left)) performs the most important role in this application.

\figthreeshader

\subsubsection{Camera and Light Control}

Secord \etal \cite{Secord11} presented a computational perceptual model for predicting the goodness of viewing directions for 3D models;
however, their model is limited to the view direction and does not consider any other factors.
We feel that good views will change according to other conditions, such as perspective and lighting.
In this scenario, we chose a camera and light control task in a simple 3D scene consisting of a 3D model, a perspective camera, and a point light.
Eight parameters are to be tweaked in total, including camera position ($x, y, z$), camera field of view, light position ($x, y, z$), and intensity of light.
We used the dragon model.
The orientation of the camera is automatically set such that it always looks at the center of the model.
We asked crowd workers to choose the better one with the same instruction.
The results indicate a highly nonlinear relationship between the camera and light parameters.
See \autoref{fig:three:shader} (Right).
When the camera comes to the left side (\ie camera.$z < 0.0$) from the right side (\ie camera.$z > 0.0$) of the dragon model, the visualization tells the user that he or she should also move the light to the left side (\ie light.$z < 0.0$) so that the model is adequately lit.

\subsubsection{Facial Expression Modeling}

Blendshape is a standard approach to control the facial expressions of virtual characters \cite{Lewis14}, where a face model has a set of predefined continuous parameters and its expression is controlled by them.
The space of ``valid'' expressions is actually quite small in most cases, which means that careful tweaking is required to ensure natural, unbroken expressions.
For this experiment, we used a head model parametrized by 53 parameters.
We suppose that the goodness in this design scenario is the validity of the facial expression, so we asked crowd workers to choose the better ``natural (unbroken)'' expression.
\autoref{fig:three:face} shows some designs and their estimated goodness values.
It is observed that the constructed goodness function successfully provides reasonable values even for this high-dimensional application.

\figthreeface


\section{Design Optimization with Crowds}
\label{sec:siggraph}

In this section, we illustrate a computational design method of optimizing visual design parameters by crowdsourced human computation \cite{Koyama17}.
This work defines a concept of \emph{crowd-powered visual design optimizer} as a system that finds an optimal design that maximizes some perceptual function from a given design space and, to enable this, bases its optimization algorithm upon the use of crowdsourced human computation.
This work offers an implementation of this concept, where the system decomposes the entire problem into a sequence of one-dimensional slider manipulation microtasks (\autoref{fig:chapter5:tas_concept} (Right)).
Crowd workers complete the tasks independently, and the system gradually reaches the optimal solution using the crowds' responses.
\autoref{fig:chapter5:approach} illustrates this concept.

\subsection{Interaction Design}

This work envisions the following scenario:
the user can push a ``Crowdsource'' button in design software for running the crowd-powered optimization process, and then he or she can obtain the ``best'' parameter set without any further interaction, as shown in \autoref{fig:chapter5:approach}.
The provided parameter set can be used, for example, as a final product or a good starting point for further tweaking.
This work investigates how to enable this function in general design applications without relying on domain-specific knowledge.

\figapproach

\subsection{Maximizing the Objective Function}
\label{chapter:5:approach}

To build a computational framework for optimizing design parameters by crowds, this work extends \emph{Bayesian optimization} techniques \cite{Shahriari16}.
Standard Bayesian optimization techniques are based on \emph{function-value} oracles;
however, with crowds and perceptual objective functions, it is not realistic to query function values as discussed in \autoref{sec:concept}.
Brochu \etal \shortcite{Brochu07} extended Bayesian optimization so that it could use pairwise-comparison oracles instead of function values;
however, the information from a pairwise comparison task is limited, and so the optimization convergence is impractically slow.
This work presents an alternative extension of Bayesian optimization based on \emph{line search} oracles instead of function-value or pairwise-comparison oracles.
The line search oracle is provided by a \emph{single-slider manipulation} query;
crowds are asked to adjust a single slider for exploring the design alternatives mapped to the slider and to return the slider value that provides the best design (see \autoref{sec:microtask} for the the discussion on this microtask design).

\subsubsection{Slider Space for Line Search}

The system lets human processors adjust a slider; that is, find a maximum in a one-dimensional continuous space.
We call this space the \emph{slider space}.
Technically, this space is not necessarily linear with respect to the target design space $\calX$ (\ie forming a straight line segment in $\calX$);
however, in this work, the case of a straight line is considered for simplicity and for the sake of not unnecessarily confusing crowd workers during the task.

At the beginning of the optimization process, the algorithm does not have any data about the target design space $\calX$ or the goodness function $g(\cdot)$.
Thus, for the initial slider space, we simply choose two random points in $\calX$ and connect them by a line segment.

For each iteration, we want to arrange the next slider space so that it is as ``meaningful'' as possible for finding $\bfx^{*}$.
We propose to construct the slider space $\calS$ such that one end is at the \emph{current-best} position $\bfx^{+}$ and the other one is at the \emph{best-expected-improving} position $\bfx^\text{EI}$.
Suppose that we have observed $t$ responses so far, and we are going to query the next oracle.
The slider space for the next iteration (\ie $\calS_{t + 1}$) is constructed by connecting
\begin{align}
\bfx^{+}_t       &= \amax{\bfx \in \{ \bfx_i \}_{i = 1}^{N_t}} \mu_t(\bfx), \\
\bfx^\text{EI}_t &= \amax{\bfx \in \calX} a^\text{EI}_t(\bfx),
\end{align}
where $\{ \bfx_i \}_{i = 1}^{N_t}$ is the set of observed data points, and $\mu_t(\cdot)$ and $a^\text{EI}_t(\cdot)$ are the predicted mean function and the \emph{acquisition function} calculated from the current data.
$\mu(\cdot)$ and $a^\text{EI}(\cdot)$ can be calculated based on an extension of \emph{Bayesian optimization} techniques;
see the original publication \cite{Koyama17} for details.

\paragraph{Example Optimization Sequence}

\autoref{fig:chapter5:slider_bayes} illustrates an example optimization sequence in which the framework is applied to a two-dimensional test function and the oracles are synthesized by a machine processor.
The process begins with a random slider space.
After several iterations, it reaches a good solution.

\figsliderbayes

\subsubsection{Implementation with Crowds}
\label{chapter:5:crowdsourcing}

Each crowd worker may respond with some ``noise,'' so averaging the responses from a sufficient number of crowd workers should provide a good approximation of the underlying common preference.
To take this into account, in each iteration, the system gathers responses from multiple crowd workers by using the same slider space, and then the system calculates the median of the provided slider tick positions and uses it for calculating $\bfx^\text{chosen}$.

\subsection{Applications}

We tested our framework in two typical parameter tweaking scenarios:
photo color enhancement and material appearance design.
All the results shown in this section were generated with 15 iterations.
For each iteration, our system deployed seven microtasks, and it proceeded to the next iteration once it had obtained at least five responses.
We paid $0.05$ USD for each microtask execution, so the total payment to the crowds was $5.25$ USD for each result.
Typically, we obtained a result in a few hours (\eg the examples in \autoref{fig:chapter5:photo_photoshop} took about $68$ minutes on average).

\subsubsection{Photo Color Enhancement}

Here, we used the same six parameters as in the previous section.
First, we compared our optimization with the auto-enhancement functions in commercial software packages.
We compared the results of our enhancement (with 15 iterations) with Adobe Photoshop CC\footnote{\url{http://www.adobe.com/products/photoshop.html}} and Adobe Photoshop Lightroom CC\footnote{\url{http://www.adobe.com/products/photoshop-lightroom.html}}.
\autoref{fig:chapter5:photo_photoshop} shows the results.
To quantify the degree of success of each enhancement, we conducted a crowdsourced study in which we asked crowd workers to identify which image looks best among the three enhancement results and the original image.
The numbers in \autoref{fig:chapter5:photo_photoshop} represent the results.
The photos enhanced by our crowd-powered optimization were preferred over the others in these cases.
These results indicate that our method can successfully produce a ``people's choice.''
This represents one of the advantages of the crowd-powered optimization.

\figphotophotoshop

Next, we repeated the same optimization procedure three times with different initial conditions (Trial A, B, and C).
\autoref{fig:chapter5:photo_free} (Left) shows the sequences of enhanced photographs over the iterations.
We measured the differences between the trials by using a \emph{perceptual color metric} based on CIEDE2000;
we measured the perceptual distance for each pixel in the enhanced photographs and calculated the mean over all the pixels.
\autoref{fig:chapter5:photo_free} (Right) shows the results.
It shows that the distances become small rapidly in the first four or five iterations, and they approach similar enhancement even though the initial conditions are quite different.

\figphotofree

\subsubsection{Material Appearance Design}

In this experiment, ``Standard Shader'' provided in Unity 5 was used as the target design space, where
the material appearance was parametrized by albedo lightness, specular lightness, and smoothness.
The number of free parameters was three in monotone and seven in full color.

The presented framework enables the automatic adjustment of shader parameters based on a user-specified reference photograph;
it minimizes the perceptual distance between the appearance in the photograph and the produced appearance by the shader.
In the microtasks, crowd workers were showed both the reference photograph and a rendered image with a slider, side by side, and asked to adjust the slider until their appearances were as similar as possible.
\autoref{fig:chapter5:teapot} shows the results for both monotone and full color spaces.

Another usage is that the user can specify textual instructions instead of reference photographs.
\autoref{fig:chapter5:teapot_instruction} illustrates the results of this usage, where crowd workers were instructed to adjust the slider so that it looks like ``brushed stainless,'' ``dark blue plastic,'' and so on.
This is not easy when a human-in-the-loop approach is not taken.

\figteapot
\figteapotinstruction


\section{Discussions}

\subsection{Usage of Computation: Estimation vs.\ Maximization}

We have considered two usages involving computational techniques.
The first approach, that is the estimation of the goodness function $g(\cdot)$, has several advantages compared to the other approach:
\begin{itemize}
\item
The user can maintain control in terms of how to explore the design space $\calX$.
That is, he or she is not forced to follow the computational guidance by the system.
\item
The chosen solution $\bfx^{*}$ is always ensured to be optimal for the target user, as the ``true'' goodness function used for deciding the final solution is owned by the user, which can be different from the ``estimated'' goodness function used for guidance.
\item
Even if the estimation is not perfect, it can still guide the user to explore the design space $\calX$ effectively.
In \autoref{sec:uist}, it was observed that the estimated goodness function was often useful for providing a good starting point for further exploration and for eliminating meaningless visits of bad designs.
\item
This approach can be seamlessly integrated in existing practical scenarios because it does not intend to replace existing workflows but does augment (or enhance) existing workflows.
\end{itemize}

In contrast, the second approach, that is the maximization of the goodness function $g(\cdot)$, has different advantages:
\begin{itemize}
\item
The user does not need to care about the strategy of how design exploration should proceed.
This could enable a new paradigm for aesthetic design and solve many constraints with respect to user experience.
For example, users are released from the need to understand and learn the effects of each design parameter.
\item
The user no longer needs to interact with the system, enabling fully automatic workflows.
This further broadens the possible usage scenarios.
\item
The found solutions by this approach can be used as either final products or good starting points for further manual refinement.
In \autoref{sec:siggraph}, it was observed that most of the results were not necessarily perfect but quite acceptable as final products.
\item
This approach aims to find the optimal solution as efficiently as possible, based on optimization techniques.
For example, the second illustrative example uses Bayesian optimization techniques so as to reduce the number of necessary iterations.
Compared to the approach of estimating the goodness function $g(\cdot)$ everywhere in the design space $\calX$, whose computational cost is exponential with respect to the dimensionality, this approach may ease this problem in high-dimensional design spaces.
\end{itemize}

A hybrid approach between these two approaches is also possible.
For example, partially estimating the goodness function $g(\cdot)$ around the expected maximum may be useful for supporting users to effectively explore the design space.
Investigating this possibility is an important future work.

\subsection{Advantages and Disadvantages of Crowds as a Data Source}

We have discussed the use of microtask-based crowdsourcing as a data source for computation.
Other possibilities for obtaining preference data include utilizing user's editing history \cite{Koyama16}, relying on activities in an online community \cite{Talton09}, and inserting an explicit training phase \cite{Kapoor14}.
Here, we summarize the advantages and disadvantages of this data source.
\begin{description}
\item[Application Domain.]
The use of microtask-based crowdsourcing limits its application to the domains where even unskilled, non-expert crowd workers can adequately assess the quality of designs.
An example of such domains is photo color enhancement;
it can be a valid assumption that most crowd workers have their preference on photo color enhancement, since enhanced photographs are ubiquitously seen in daily lives (\eg in product advertisements).
However, some design domains exist where only experts can adequately assess the quality of designs;
in such cases, asking crowds may not be a reasonable choice.
\item[Whose Preference?]
Asking many people is advantageous in that the system can reliably understand ``people's choice'' and utilize it.
The illustrative examples described in the previous sections assume the existence of a ``general'' (or universal) preference shared among crowds and ask multiple workers to perform microtasks to observe the general preference.
However, in some design domains, crowds from different backgrounds can have clearly different preferences (\cf \cite{Reinecke14}).
In such cases, it is necessary to take each worker's preference into account in the computation;
otherwise, the computed preference would be less meaningful.
It is also notable that the user's personal preference is not reflected in computation when crowdsourcing is the only source of data;
to incorporate the user's preference, other data sources such as the user's editing history have to be used in combination with crowdsouring.
\item[Data Gathering.]
Microtask-based crowdsourcing enables on-demand and stable generation of new data, which is a large advantage of this approach.
One of the limitations in this approach is the latency for data generation;
although there are techniques for enabling real-time crowdsourcing (\eg \cite{Bernstein11}), it is generally difficult to obtain necessary responses interactively.
Another limitation is that crowds may provide poor-quality data due to mistaking or cheating, which requires to use computational techniques that are robust for such ``noisy'' data.
The monetary cost is also an inevitable limitation.
\end{description}

\subsection{Remaining Challenges}

We have discussed computational design methods under many assumptions.
To ease the assumptions and develop more practical methods, a number of possible challenges remain as next steps.

\paragraph{Design Space Parametrization.}

We have assumed that the target design space is appropriately parametrized in advance.
It is important to seek appropriate parametrization techniques to broaden the applications of crowd-powered computational design.
For example, it is difficult to directly handle very high-dimensional design spaces.
We consider that dimensionality reduction techniques could be helpful for many problems with high-dimensional design spaces.
However, this is still non-trivial because, unlike typical problems in data science, the resulting space in this case has to be either designer-friendly or optimization-friendly (or both) for maximizing aesthetic preference.
Recently, Yumer \etal \shortcite{Yumer15UIST} showed that autoencoder networks can be used for converting a high-dimensional, visually discontinuous design space to a lower-dimensional, visually continuous design space that is more desirable for design exploration.
Incorporating human preference in dimensionality reduction of design spaces is an interesting future work.

\paragraph{Discrete Parameters.}

We focused on continuous parameters, and did not discuss how to handle discrete design parameters, such as fonts \cite{ODonovan14TOG} and web design templates \cite{Chaudhuri13}.
The remaining challenges to handle such discrete parameters include how to represent goodness functions for design spaces including discrete parameters and how to facilitate users' interactive explorations.
Investigating techniques for jointly handling discrete and continuous parameters is another potential future work.

\paragraph{Locally Optimal Design Alternatives.}

In some scenarios, totally different design alternatives can be equally ``best,'' and it can be hard to determine which is better.
For example, in Adobe Color CC\footnote{\url{https://color.adobe.com/}}, which is a user community platform to make, explore, and share \emph{color palettes} (a set of colors usually consisting of five colors), there are a number of (almost) equally popular color palettes that have been preferred by many users, as shown in \autoref{fig:chapter6:colors}.
In this case, if we assume the existence of a goodness function for color palettes, the popular palettes can be considered as local maximums of the goodness function.
Considering that the goal is to support design activities, it may not be effective to assume that there is a sole global maximum in this design space and to guide the user toward the single maximum;
rather, it may be more desirable to provide a variety of good design alternatives.
There is a room for investigation about how computation can support such design scenarios.

\figsixcolors

\paragraph{Evaluation Methodology.}

One of the issues in computational design driven by aesthetic preference is the lack of an established, general methodology of evaluating each new method.
Validation of methods in this domain is challenging for several reasons.
The first reason is the difficulty of defining ``correct'' aesthetic preference, which can be highly dependent on scenarios.
Also, as the ultimate goal is the support of design activities, the effectiveness needs to be evaluated by designers.
Methods in this domain are built on many assumptions, each of which is difficult to validate.
We consider that an important future work would be to establish general evaluation schemes.

\paragraph{More Sophisticated Models of Crowd Behaviors.}

The two examples described in the previous sections were built on an assumption of crowd workers:
crowd workers share a common goodness function, and each crowd worker responds based on the function with some noise.
Thus, it is assumed that the common goodness function is observed by asking many crowd workers and then averaging their responses.
This assumption may be valid in some scenarios, but may not be in many other scenarios;
for example, crowds may form several clusters with respect to their aesthetic preference.
Modeling such more complex properties of crowds is an important future challenge.

\paragraph{Incorporating Domain-Specific Heuristics.}

We have tried to use minimal domain-specific knowledge so that the discussions made in this chapter are as generally applicable as possible.
However, it is possible to make computational design methods more practical for certain specific scenarios by making full use of domain-specific heuristics.
For example, if one builds a software program to tweak the viewpoints of 3D objects, the heuristic features and the pre-trained model in \cite{Secord11} could be jointly used with the methods described in the previous sections.

\paragraph{Combining General and Personal Preference.}

We discussed how to handle crowds' \emph{general} preference.
However, it would also be beneficial if we could handle users' \emph{personal} preference;
Koyama \etal \shortcite{Koyama16} presented a method for learning personal preference to facilitate design exploration and reported that this approach was appreciated by professional designers.
Both approaches have advantages and disadvantages;
to complement the disadvantages of each approach, we envision that the combination of these two approaches may be useful and worth investigating.

\subsection{Future Directions}

Finally, we conclude this chapter with discussions of several future research directions on computational design methods powered by crowds.

\subsubsection{Beyond Parametric Design}

We focused on parametric design, where the design space is reasonably parametrized beforehand.
This is often performed for the \emph{refinement} of a design.
On the other hand, the \emph{generation} of a design from scratch is also an important step of an entire design process.
It is an open question whether we can handle such free-form designs as an extension of parameter tweaking.

Some class of design generation processes can be described as a sequence of executable commands (or operations).
In this case, the design space can be modeled as a tree structure whose leaves and edges represent visual designs and executable commands, respectively, and the design goal can be formulated to find the best leaf node from this tree.
For this, tree search optimization techniques (\eg the branch and bound method) might be useful.
It is also notable that interactive evolutionary computation (IEC) can be used to generate designs while accounting for human evaluation in the loop.
For example, Sims \shortcite{Sims91} showed that IEC can generate interesting designs beyond predefined parametric design spaces.

\subsubsection{More Complex Design Objectives}

In practical design scenarios, designers may have to solve complex problems with more than one design objective.
For example, when a graphic designer designs an advertisement targeted at mainly women, he or she has to bias the objective toward women's aesthetic preference.
In this case, it is possible to formulate the design problem as
\begin{align}
\bfx^{*} = \amax{\bfx \in \calX} \{ w_\text{male} g_\text{male}(\bfx) + w_\text{female} g_\text{female}(\bfx) \},
\end{align}
where $g_\text{male}(\cdot)$ and $g_\text{female}(\cdot)$ are the goodness functions owned by men and women, respectively, and $w_\text{male}$ and $w_\text{female}$ are the weights for adjusting the bias, which can be $w_\text{male} < w_\text{female}$ in this case.
With crowdsourced human computation, this could be solved by utilizing the demographic information of crowd workers (\cf \cite{Reinecke14}).
Another complex scenario is a case in which some additional design criteria are expected to be at least satisfied, but do not have to be maximized.
For example, a client may want a design that is as preferred by young people as possible and at the same time is ``acceptable'' by elderly people.
In this case, the problem can be formulated as a constrained optimization.
Under these complex conditions, it should be difficult for designers to manually explore designs.
We believe that this is the part that computational techniques need to facilitate.

\subsubsection{Computational Creative Design with Crowds}

We have considered aesthetic preference as the target criterion in design activities.
Another important aspect of design is \emph{creativity}.
One of the keys to provide creative inspirations to designers is \emph{unexpectedness} \cite{CohenOr16}.
Some researchers have interpreted such unexpectedness as \emph{diversity} in design alternatives and have explicitly formulated optimization problems for finding the most diverse set of design alternatives (\eg \cite{Won14}).
We believe that there are many interesting opportunities to investigate computational frameworks for achieving creative designs by making use of crowds' creativity.


\section{Summary}

In this chapter, we discussed the possible mechanisms, illustrative examples, and future challenges of computational design methods with crowds.
Especially, we focused on the facilitation of parametric design (\ie parameter tweaking) and then formulated the design process as a numerical optimization problem, where the objective function to be maximized was based on perceptual preference.
We illustrated the ideas of using crowdsourced human computation for this problem, either for the estimation of the objective function or for the maximization of the objective function.

\bibliographystyle{plainnat-doi}
\bibliography{main}

\begin{thebibliography}{53}
\providecommand{\natexlab}[1]{#1}
\providecommand{\doi}[1]{doi: \url{https://doi.org/#1}}

\bibitem[Allen et~al.(2003)Allen, Curless, and Popovi\'{c}]{Allen03}
Brett Allen, Brian Curless, and Zoran Popovi\'{c}.
\newblock The space of human body shapes: Reconstruction and parameterization
  from range scans.
\newblock \emph{ACM Trans. Graph.}, 22\penalty0 (3):\penalty0 587--594, July
  2003.
\newblock \doi{10.1145/882262.882311}.

\bibitem[Bailly et~al.(2013)Bailly, Oulasvirta, K\"{o}tzing, and
  Hoppe]{Bailly13}
Gilles Bailly, Antti Oulasvirta, Timo K\"{o}tzing, and Sabrina Hoppe.
\newblock {MenuOptimizer}: Interactive optimization of menu systems.
\newblock In \emph{Proceedings of the 26th Annual ACM Symposium on User
  Interface Software and Technology}, UIST '13, pages 331--342, 2013.
\newblock \doi{10.1145/2501988.2502024}.

\bibitem[Bell et~al.(2013)Bell, Upchurch, Snavely, and Bala]{Bell13}
Sean Bell, Paul Upchurch, Noah Snavely, and Kavita Bala.
\newblock Opensurfaces: A richly annotated catalog of surface appearance.
\newblock \emph{ACM Trans. Graph.}, 32\penalty0 (4):\penalty0 111:1--111:17,
  July 2013.
\newblock \doi{10.1145/2461912.2462002}.
\newblock URL: \url{http://doi.acm.org/10.1145/2461912.2462002}.

\bibitem[Bernstein et~al.(2010)Bernstein, Little, Miller, Hartmann, Ackerman,
  Karger, Crowell, and Panovich]{Bernstein10}
Michael~S. Bernstein, Greg Little, Robert~C. Miller, Bj\"{o}rn Hartmann,
  Mark~S. Ackerman, David~R. Karger, David Crowell, and Katrina Panovich.
\newblock Soylent: A word processor with a crowd inside.
\newblock In \emph{Proceedings of the 23rd Annual ACM Symposium on User
  Interface Software and Technology}, UIST '10, pages 313--322, 2010.
\newblock \doi{10.1145/1866029.1866078}.

\bibitem[Bernstein et~al.(2011)Bernstein, Brandt, Miller, and
  Karger]{Bernstein11}
Michael~S. Bernstein, Joel Brandt, Robert~C. Miller, and David~R. Karger.
\newblock Crowds in two seconds: Enabling realtime crowd-powered interfaces.
\newblock In \emph{Proceedings of the 24th Annual ACM Symposium on User
  Interface Software and Technology}, UIST '11, pages 33--42, 2011.
\newblock \doi{10.1145/2047196.2047201}.
\newblock URL: \url{http://doi.acm.org/10.1145/2047196.2047201}.

\bibitem[Bishop(1995)]{Bishop95}
Christopher~M. Bishop.
\newblock \emph{Neural Networks for Pattern Recognition}.
\newblock Oxford University Press, Inc., 1995.

\bibitem[Blanz and Vetter(1999)]{Blanz99}
Volker Blanz and Thomas Vetter.
\newblock A morphable model for the synthesis of 3d faces.
\newblock In \emph{Proceedings of the 26th Annual Conference on Computer
  Graphics and Interactive Techniques}, SIGGRAPH '99, pages 187--194, 1999.
\newblock \doi{10.1145/311535.311556}.

\bibitem[Brochu et~al.(2007)Brochu, de~Freitas, and Ghosh]{Brochu07}
Eric Brochu, Nando de~Freitas, and Abhijeet Ghosh.
\newblock Active preference learning with discrete choice data.
\newblock In \emph{Advances in Neural Information Processing Systems 20}, NIPS
  '07, pages 409--416, 2007.
\newblock URL:
  \url{http://papers.nips.cc/paper/3219-active-preference-learning-with-discrete-choice-data.pdf}.

\bibitem[Bychkovsky et~al.(2011)Bychkovsky, Paris, Chan, and
  Durand]{Bychkovsky11}
Vladimir Bychkovsky, Sylvain Paris, Eric Chan, and Fr{\'e}do Durand.
\newblock Learning photographic global tonal adjustment with a database of
  input/output image pairs.
\newblock In \emph{Proceedings of the 24th IEEE Conference on Computer Vision
  and Pattern Recognition}, CVPR '11, pages 97--104, 2011.
\newblock \doi{10.1109/CVPR.2011.5995413}.

\bibitem[Chaudhuri et~al.(2013)Chaudhuri, Kalogerakis, Giguere, and
  Funkhouser]{Chaudhuri13}
Siddhartha Chaudhuri, Evangelos Kalogerakis, Stephen Giguere, and Thomas
  Funkhouser.
\newblock Attribit: Content creation with semantic attributes.
\newblock In \emph{Proceedings of the 26th Annual ACM Symposium on User
  Interface Software and Technology}, UIST '13, pages 193--202, 2013.
\newblock \doi{10.1145/2501988.2502008}.

\bibitem[Cohen-Or and Zhang(2016)]{CohenOr16}
Daniel Cohen-Or and Hao Zhang.
\newblock From inspired modeling to creative modeling.
\newblock \emph{The Visual Computer}, 32\penalty0 (1):\penalty0 7--14, 2016.
\newblock \doi{10.1007/s00371-015-1193-9}.

\bibitem[Gajos and Weld(2004)]{Gajos04}
Krzysztof Gajos and Daniel~S. Weld.
\newblock {SUPPLE}: Automatically generating user interfaces.
\newblock In \emph{Proceedings of the 9th International Conference on
  Intelligent User Interfaces}, IUI '04, pages 93--100, 2004.
\newblock \doi{10.1145/964442.964461}.

\bibitem[Gingold et~al.(2012)Gingold, Shamir, and Cohen-Or]{Gingold12}
Yotam Gingold, Ariel Shamir, and Daniel Cohen-Or.
\newblock Micro perceptual human computation for visual tasks.
\newblock \emph{ACM Trans. Graph.}, 31\penalty0 (5):\penalty0 119:1--119:12,
  September 2012.
\newblock \doi{10.1145/2231816.2231817}.

\bibitem[Howe(2006{\natexlab{a}})]{Howe06}
Jeff Howe.
\newblock The rise of crowdsourcing.
\newblock \url{https://www.wired.com/2006/06/crowds/}, 2006{\natexlab{a}}.
\newblock Last checked: October 23, 2016.

\bibitem[Howe(2006{\natexlab{b}})]{Howe06b}
Jeff Howe.
\newblock Crowdsourcing: A definition.
\newblock
  \url{http://crowdsourcing.typepad.com/cs/2006/06/crowdsourcing_a.html},
  2006{\natexlab{b}}.
\newblock Last checked: October 23, 2016.

\bibitem[Huber et~al.(2017)Huber, Reinecke, and Gajos]{Huber17}
Bernd Huber, Katharina Reinecke, and Krzysztof~Z. Gajos.
\newblock The effect of performance feedback on social media sharing at
  volunteer-based online experiment platforms.
\newblock In \emph{Proceedings of the 2017 CHI Conference on Human Factors in
  Computing Systems}, CHI '17, pages 1882--1886, 2017.
\newblock \doi{10.1145/3025453.3025553}.

\bibitem[Ipeirotis et~al.(2010)Ipeirotis, Provost, and Wang]{Ipeirotis10}
Panagiotis~G. Ipeirotis, Foster Provost, and Jing Wang.
\newblock Quality management on {Amazon Mechanical Turk}.
\newblock In \emph{Proceedings of the ACM SIGKDD Workshop on Human
  Computation}, HCOMP '10, pages 64--67, 2010.
\newblock \doi{10.1145/1837885.1837906}.

\bibitem[Kapoor et~al.(2014)Kapoor, Caicedo, Lischinski, and Kang]{Kapoor14}
Ashish Kapoor, Juan~C. Caicedo, Dani Lischinski, and Sing~Bing Kang.
\newblock Collaborative personalization of image enhancement.
\newblock \emph{International Journal of Computer Vision}, 108\penalty0
  (1):\penalty0 148--164, 2014.
\newblock \doi{10.1007/s11263-013-0675-3}.

\bibitem[Kittur et~al.(2008)Kittur, Chi, and Suh]{Kittur08}
Aniket Kittur, Ed~H. Chi, and Bongwon Suh.
\newblock Crowdsourcing user studies with {Mechanical Turk}.
\newblock In \emph{Proceedings of the SIGCHI Conference on Human Factors in
  Computing Systems}, CHI '08, pages 453--456, 2008.
\newblock \doi{10.1145/1357054.1357127}.

\bibitem[Koyama et~al.(2014)Koyama, Sakamoto, and Igarashi]{Koyama14}
Yuki Koyama, Daisuke Sakamoto, and Takeo Igarashi.
\newblock Crowd-powered parameter analysis for visual design exploration.
\newblock In \emph{Proceedings of the 27th Annual ACM Symposium on User
  Interface Software and Technology}, UIST '14, pages 65--74, 2014.
\newblock \doi{10.1145/2642918.2647386}.

\bibitem[Koyama et~al.(2015)Koyama, Sueda, Steinhardt, Igarashi, Shamir, and
  Matusik]{Koyama15}
Yuki Koyama, Shinjiro Sueda, Emma Steinhardt, Takeo Igarashi, Ariel Shamir, and
  Wojciech Matusik.
\newblock {AutoConnect}: Computational design of 3d-printable connectors.
\newblock \emph{ACM Trans. Graph.}, 34\penalty0 (6):\penalty0 231:1--231:11,
  October 2015.
\newblock \doi{10.1145/2816795.2818060}.

\bibitem[Koyama et~al.(2016)Koyama, Sakamoto, and Igarashi]{Koyama16}
Yuki Koyama, Daisuke Sakamoto, and Takeo Igarashi.
\newblock {SelPh}: Progressive learning and support of manual photo color
  enhancement.
\newblock In \emph{Proceedings of the 2016 CHI Conference on Human Factors in
  Computing Systems}, CHI '16, pages 2520--2532, 2016.
\newblock \doi{10.1145/2858036.2858111}.

\bibitem[Koyama et~al.(2017)Koyama, Sato, Sakamoto, and Igarashi]{Koyama17}
Yuki Koyama, Issei Sato, Daisuke Sakamoto, and Takeo Igarashi.
\newblock Sequential line search for efficient visual design optimization by
  crowds.
\newblock \emph{ACM Trans. Graph.}, 36\penalty0 (4):\penalty0 48:1--48:11, July
  2017.
\newblock \doi{10.1145/3072959.3073598}.

\bibitem[Lewis et~al.(2014)Lewis, Anjyo, Rhee, Zhang, Pighin, and
  Deng]{Lewis14}
J.~P. Lewis, Ken Anjyo, Taehyun Rhee, Mengjie Zhang, Fred Pighin, and Zhigang
  Deng.
\newblock Practice and theory of blendshape facial models.
\newblock In \emph{Eurographics 2014---State of the Art Reports}, Eurographics
  '14, pages 199--218, 2014.
\newblock \doi{10.2312/egst.20141042}.

\bibitem[Little et~al.(2010)Little, Chilton, Goldman, and Miller]{Little10}
Greg Little, Lydia~B. Chilton, Max Goldman, and Robert~C. Miller.
\newblock Turkit: Human computation algorithms on mechanical turk.
\newblock In \emph{Proceedings of the 23rd Annual ACM Symposium on User
  Interface Software and Technology}, UIST '10, pages 57--66, 2010.
\newblock \doi{10.1145/1866029.1866040}.

\bibitem[Matusik et~al.(2003)Matusik, Pfister, Brand, and McMillan]{Matusik03}
Wojciech Matusik, Hanspeter Pfister, Matt Brand, and Leonard McMillan.
\newblock A data-driven reflectance model.
\newblock \emph{ACM Trans. Graph.}, 22\penalty0 (3):\penalty0 759--769, July
  2003.
\newblock \doi{10.1145/882262.882343}.

\bibitem[Miniukovich and De~Angeli(2015)]{Miniukovich15}
Aliaksei Miniukovich and Antonella De~Angeli.
\newblock Computation of interface aesthetics.
\newblock In \emph{Proceedings of the 33rd Annual ACM Conference on Human
  Factors in Computing Systems}, CHI '15, pages 1163--1172, 2015.
\newblock \doi{10.1145/2702123.2702575}.

\bibitem[Morishima et~al.(2014)Morishima, Amer-Yahia, and Roy]{Morishima14}
Atsuyuki Morishima, Sihem Amer-Yahia, and Senjuti~Basu Roy.
\newblock {Crowd4U}: An initiative for constructing an open academic
  crowdsourcing network.
\newblock In \emph{Proceedings of Second AAAI Conference on Human Computation
  and Crowdsourcing -- Works in Progress Abstracts}, HCOMP '14, 2014.
\newblock URL:
  \url{https://www.aaai.org/ocs/index.php/HCOMP/HCOMP14/paper/view/9061}.

\bibitem[Nielsen et~al.(2015)Nielsen, Jensen, and Ramamoorthi]{Nielsen15}
Jannik~Boll Nielsen, Henrik~Wann Jensen, and Ravi Ramamoorthi.
\newblock On optimal, minimal {BRDF} sampling for reflectance acquisition.
\newblock \emph{ACM Trans. Graph.}, 34\penalty0 (6):\penalty0 186:1--186:11,
  October 2015.
\newblock \doi{10.1145/2816795.2818085}.

\bibitem[O'Donovan et~al.(2011)O'Donovan, Agarwala, and Hertzmann]{ODonovan11}
Peter O'Donovan, Aseem Agarwala, and Aaron Hertzmann.
\newblock Color compatibility from large datasets.
\newblock \emph{ACM Trans. Graph.}, 30\penalty0 (4):\penalty0 63:1--63:12, July
  2011.
\newblock \doi{10.1145/2010324.1964958}.

\bibitem[O'Donovan et~al.(2014{\natexlab{a}})O'Donovan, Agarwala, and
  Hertzmann]{ODonovan14TVCG}
Peter O'Donovan, Aseem Agarwala, and Aaron Hertzmann.
\newblock Learning layouts for single-page graphic designs.
\newblock \emph{IEEE Transactions on Visualization and Computer Graphics},
  20\penalty0 (8):\penalty0 1200--1213, August 2014{\natexlab{a}}.
\newblock \doi{10.1109/TVCG.2014.48}.

\bibitem[O'Donovan et~al.(2014{\natexlab{b}})O'Donovan, L\={\i}beks, Agarwala,
  and Hertzmann]{ODonovan14TOG}
Peter O'Donovan, J\={a}nis L\={\i}beks, Aseem Agarwala, and Aaron Hertzmann.
\newblock Exploratory font selection using crowdsourced attributes.
\newblock \emph{ACM Trans. Graph.}, 33\penalty0 (4):\penalty0 92:1--92:9, July
  2014{\natexlab{b}}.
\newblock \doi{10.1145/2601097.2601110}.

\bibitem[Oulasvirta and Karrenbauer(2018)]{Oulasvirta18}
Antti Oulasvirta and Andreas Karrenbauer.
\newblock Combinatorial optimization for user interface design.
\newblock In Antti Oulasvirta, Per~Ola Kristensson, Xiaojun Bi, and Andrew
  Howes (Eds.), \emph{Computational Interaction}, chapter~4, pages 97--120.
  Oxford University Press, 2018.
\newblock \doi{10.1093/oso/9780198799603.001.0001}.

\bibitem[Pr{\'e}vost et~al.(2013)Pr{\'e}vost, Whiting, Lefebvre, and
  Sorkine-Hornung]{Prevost13}
Romain Pr{\'e}vost, Emily Whiting, Sylvain Lefebvre, and Olga Sorkine-Hornung.
\newblock Make it stand: Balancing shapes for 3d fabrication.
\newblock \emph{ACM Trans. Graph.}, 32\penalty0 (4):\penalty0 81:1--81:10, July
  2013.
\newblock \doi{10.1145/2461912.2461957}.

\bibitem[Quinn and Bederson(2011)]{Quinn11}
Alexander~J. Quinn and Benjamin~B. Bederson.
\newblock Human computation: A survey and taxonomy of a growing field.
\newblock In \emph{Proceedings of the SIGCHI Conference on Human Factors in
  Computing Systems}, CHI '11, pages 1403--1412, 2011.
\newblock \doi{10.1145/1978942.1979148}.

\bibitem[Reinecke and Gajos(2014)]{Reinecke14}
Katharina Reinecke and Krzysztof~Z. Gajos.
\newblock Quantifying visual preferences around the world.
\newblock In \emph{Proceedings of the SIGCHI Conference on Human Factors in
  Computing Systems}, CHI '14, pages 11--20, 2014.
\newblock \doi{10.1145/2556288.2557052}.

\bibitem[Secord et~al.(2011)Secord, Lu, Finkelstein, Singh, and
  Nealen]{Secord11}
Adrian Secord, Jingwan Lu, Adam Finkelstein, Manish Singh, and Andrew Nealen.
\newblock Perceptual models of viewpoint preference.
\newblock \emph{ACM Trans. Graph.}, 30\penalty0 (5):\penalty0 109:1--109:12,
  October 2011.
\newblock \doi{10.1145/2019627.2019628}.

\bibitem[Shahriari et~al.(2016)Shahriari, Swersky, Wang, Adams, and
  de~Freitas]{Shahriari16}
Bobak Shahriari, Kevin Swersky, Ziyu Wang, Ryan~P. Adams, and Nando de~Freitas.
\newblock Taking the human out of the loop: A review of {Bayesian}
  optimization.
\newblock \emph{Proceedings of the IEEE}, 104\penalty0 (1):\penalty0 148--175,
  January 2016.
\newblock \doi{10.1109/JPROC.2015.2494218}.

\bibitem[Sigal et~al.(2015)Sigal, Mahler, Diaz, McIntosh, Carter, Richards, and
  Hodgins]{Sigal15}
Leonid Sigal, Moshe Mahler, Spencer Diaz, Kyna McIntosh, Elizabeth Carter,
  Timothy Richards, and Jessica Hodgins.
\newblock A perceptual control space for garment simulation.
\newblock \emph{ACM Trans. Graph.}, 34\penalty0 (4):\penalty0 117:1--117:10,
  July 2015.
\newblock \doi{10.1145/2766971}.

\bibitem[Sims(1991)]{Sims91}
Karl Sims.
\newblock Artificial evolution for computer graphics.
\newblock \emph{SIGGRAPH Comput. Graph.}, 25\penalty0 (4):\penalty0 319--328,
  July 1991.
\newblock \doi{10.1145/127719.122752}.

\bibitem[Stava et~al.(2012)Stava, Vanek, Benes, Carr, and M\v{e}ch]{Stava12}
Ondrej Stava, Juraj Vanek, Bedrich Benes, Nathan Carr, and Radom\'{\i}r
  M\v{e}ch.
\newblock Stress relief: Improving structural strength of {3D} printable
  objects.
\newblock \emph{ACM Trans. Graph.}, 31\penalty0 (4):\penalty0 48:1--48:11, July
  2012.
\newblock \doi{10.1145/2185520.2185544}.

\bibitem[Streuber et~al.(2016)Streuber, Quiros-Ramirez, Hill, Hahn, Zuffi,
  O'Toole, and Black]{Streuber16}
Stephan Streuber, M.~Alejandra Quiros-Ramirez, Matthew~Q. Hill, Carina~A. Hahn,
  Silvia Zuffi, Alice O'Toole, and Michael~J. Black.
\newblock {Body Talk}: Crowdshaping realistic 3d avatars with words.
\newblock \emph{ACM Trans. Graph.}, 35\penalty0 (4):\penalty0 54:1--54:14, July
  2016.
\newblock \doi{10.1145/2897824.2925981}.

\bibitem[Talton et~al.(2009)Talton, Gibson, Yang, Hanrahan, and
  Koltun]{Talton09}
Jerry~O. Talton, Daniel Gibson, Lingfeng Yang, Pat Hanrahan, and Vladlen
  Koltun.
\newblock Exploratory modeling with collaborative design spaces.
\newblock \emph{ACM Trans. Graph.}, 28\penalty0 (5):\penalty0 167:1--167:10,
  December 2009.
\newblock \doi{10.1145/1618452.1618513}.

\bibitem[Todi et~al.(2016)Todi, Weir, and Oulasvirta]{Todi16}
Kashyap Todi, Daryl Weir, and Antti Oulasvirta.
\newblock {Sketchplore}: Sketch and explore with a layout optimiser.
\newblock In \emph{Proceedings of the 2016 ACM Conference on Designing
  Interactive Systems}, DIS '16, pages 543--555, 2016.
\newblock \doi{10.1145/2901790.2901817}.

\bibitem[Umetani et~al.(2014)Umetani, Koyama, Schmidt, and Igarashi]{Umetani14}
Nobuyuki Umetani, Yuki Koyama, Ryan Schmidt, and Takeo Igarashi.
\newblock Pteromys: Interactive design and optimization of free-formed
  free-flight model airplanes.
\newblock \emph{ACM Trans. Graph.}, 33\penalty0 (4):\penalty0 65:1--65:10, July
  2014.
\newblock \doi{10.1145/2601097.2601129}.

\bibitem[von Ahn(2005)]{VonAhn05}
Luis von Ahn.
\newblock \emph{Human Computation}.
\newblock PhD thesis, Carnegie Mellon University, Pittsburgh, PA, USA, 2005.
\newblock URL:
  \url{http://reports-archive.adm.cs.cmu.edu/anon/2005/CMU-CS-05-193.pdf}.
\newblock CMU-CS-05-193.

\bibitem[von Ahn and Dabbish(2004)]{vonAhn04}
Luis von Ahn and Laura Dabbish.
\newblock Labeling images with a computer game.
\newblock In \emph{Proceedings of the SIGCHI Conference on Human Factors in
  Computing Systems}, CHI '04, pages 319--326, 2004.
\newblock \doi{10.1145/985692.985733}.

\bibitem[von Ahn and Dabbish(2008)]{vonAhn08ACMComm}
Luis von Ahn and Laura Dabbish.
\newblock Designing games with a purpose.
\newblock \emph{Commun. ACM}, 51\penalty0 (8):\penalty0 58--67, August 2008.
\newblock \doi{10.1145/1378704.1378719}.

\bibitem[von Ahn et~al.(2008)von Ahn, Maurer, McMillen, Abraham, and
  Blum]{vonAhn08Science}
Luis von Ahn, Benjamin Maurer, Colin McMillen, David Abraham, and Manuel Blum.
\newblock {reCAPTCHA}: Human-based character recognition via web security
  measures.
\newblock \emph{Science}, 321\penalty0 (5895):\penalty0 1465--1468, 2008.
\newblock \doi{10.1126/science.1160379}.

\bibitem[Weber and Penn(1995)]{Weber95}
Jason Weber and Joseph Penn.
\newblock Creation and rendering of realistic trees.
\newblock In \emph{Proceedings of the 22nd Annual Conference on Computer
  Graphics and Interactive Techniques}, SIGGRAPH '95, pages 119--128, 1995.
\newblock \doi{10.1145/218380.218427}.

\bibitem[Won et~al.(2014)Won, Lee, O'Sullivan, Hodgins, and Lee]{Won14}
Jungdam Won, Kyungho Lee, Carol O'Sullivan, Jessica~K. Hodgins, and Jehee Lee.
\newblock Generating and ranking diverse multi-character interactions.
\newblock \emph{ACM Trans. Graph.}, 33\penalty0 (6):\penalty0 219:1--219:12,
  November 2014.
\newblock \doi{10.1145/2661229.2661271}.

\bibitem[Yumer et~al.(2015{\natexlab{a}})Yumer, Asente, Mech, and
  Kara]{Yumer15UIST}
Mehmet~Ersin Yumer, Paul Asente, Radomir Mech, and Levent~Burak Kara.
\newblock Procedural modeling using autoencoder networks.
\newblock In \emph{Proceedings of the 28th Annual ACM Symposium on User
  Interface Software and Technology}, UIST '15, pages 109--118,
  2015{\natexlab{a}}.
\newblock \doi{10.1145/2807442.2807448}.

\bibitem[Yumer et~al.(2015{\natexlab{b}})Yumer, Chaudhuri, Hodgins, and
  Kara]{Yumer15TOG}
Mehmet~Ersin Yumer, Siddhartha Chaudhuri, Jessica~K. Hodgins, and Levent~Burak
  Kara.
\newblock Semantic shape editing using deformation handles.
\newblock \emph{ACM Trans. Graph.}, 34\penalty0 (4):\penalty0 86:1--86:12, July
  2015{\natexlab{b}}.
\newblock \doi{10.1145/2766908}.

\end{thebibliography}

\end{document}